\documentclass[useAMS,usenatbib,usegraphicx]{mn2e}
\usepackage{url,times,makeidx,xcolor,amsmath,amssymb,graphicx}

\graphicspath{{img/}}
\DeclareGraphicsExtensions{.pdf,.jpeg,.png,.eps,.tiff}

\newcommand\nfull{\ensuremath{N_\mathrm{filled}}}
\newcommand\nempty{\ensuremath{N_\mathrm{empty}}}
\newcommand\emrate{\ensuremath{r_\mathrm{r}}}
\newcommand\caprate{\ensuremath{r_\mathrm{c}}}
\newcommand\pfull{\ensuremath{p_\mathrm{filled}}}
\newcommand\pinf{\ensuremath{p_\mathrm{filled\,\infty}}}
\newcommand\pempty{\ensuremath{p_\mathrm{empty}}}
\newcommand\prel{\ensuremath{p_\mathrm{r}}}
\newcommand\pcap{\ensuremath{p_\mathrm{c}}}
\newcommand\taurel{\ensuremath{\tau_\mathrm{r}}}
\newcommand\taucap{\ensuremath{\tau_\mathrm{c}}}
\newcommand\edens{\ensuremath{n_\mathrm{e}}}
\newcommand\vect[1]{\ensuremath{\mathbf{#1}}}
\newcommand\electron{\ensuremath{\mathrm{e}^{-}}}
\newcommand\nsat{\ensuremath{n_\mathrm{sat}}}
\newcommand\ssat{\ensuremath{S_\mathrm{sat}}}
\newcommand\thetasim{\ensuremath{\theta_\mathrm{sim}}}
\newcommand\thetaexp{\ensuremath{\theta_\mathrm{exp}}}
\newcommand\trapen{\ensuremath{E_\mathrm{t}}}
\newcommand\trapsigma{\ensuremath{\sigma_\mathrm{t}}}
\newcommand\trapdens{\ensuremath{n_\mathrm{t}}}
\newcommand\dwelltime{\ensuremath{\Delta_\mathrm{t}}}

\newcommand\gaia{\textit{Gaia}}
\newcommand\hst{\textit{HST}}

\makeindex
\begin{document}
%
\title[Electrode level model of radiation damage effects on CCDs]{Electrode level Monte Carlo model of radiation damage effects on astronomical CCDs}
\author[T.\ Prod'homme,
        A.G.A.\ Brown,
        L.\ Lindegren,
        A.D.T.\ Short,
        S.W.\ Brown]{
        T.\ Prod'homme$^{1}$ \thanks{E-mail:prodhomme@strw.leidenuniv.nl},
        A.G.A.\ Brown$^{1}$, L.\ Lindegren$^{2}$, A.D.T.\ Short$^{3}$, and S.W.\ Brown$^{4}$\\
$^{1}$Leiden Observatory, Leiden University, P.O.\ Box 9513, 2300 RA, Leiden, The Netherlands\\
$^{2}$Lund Observatory, Lund University, Box 43, 22100 Lund, Sweden\\
$^{3}$ESA, ESTEC, Postbus 299, 2200 AG Noordwijk, The Netherlands\\
$^{4}$Institute of Astronomy, Madingley Road, Cambridge CB3 0HA, United Kingdom}

\maketitle

\begin{abstract}

Current optical space telescopes rely upon silicon Charge Coupled Devices (CCDs) to detect and image the incoming photons. The performance of a CCD detector depends on its ability to transfer electrons through the silicon efficiently, so that the signal from every pixel may be read out through a single amplifier. This process of electron transfer is highly susceptible to the effects of solar proton damage (or non-ionizing radiation damage). This is because charged particles passing through the CCD displace silicon atoms, introducing energy levels into the semi-conductor bandgap which act as localized electron traps. The reduction in Charge Transfer Efficiency (CTE) leads to signal loss and image smearing. The European Space Agency's astrometric {\gaia} mission will make extensive use of CCDs to create the most complete and accurate stereoscopic map to date of the Milky Way. In the context of the {\gaia} mission CTE is referred to with the complementary quantity Charge Transfer Inefficiency (CTI $= 1-$CTE). CTI is an extremely important issue that threatens {\gaia}'s performances: the CCDs are very large so that the electrons need to be transferred a long way; the focal plane is also very large and difficult to shield; the mission will operate at L2 where the direct solar protons are highly energetic (penetrating); and the science requirements on image quality are very stringent. In order to tackle this issue, in depth experimental studies and modelling efforts are being conducted to explore the possible consequences and to mitigate the anticipated effects of radiation damage. We present here a detailed Monte Carlo model which has been developed to simulate the operation of a damaged CCD at the pixel electrode level. This model implements a new approach to both the charge density distribution within a pixel and the charge capture and release probabilities, which allows the reproduction of CTI effects on a variety of measurements for a large signal level range in particular for signals of the order of a few electrons.

\end{abstract}


\section{Introduction}

We present a detailed physical Monte Carlo model of Charge Transfer
Inefficiency (CTI) caused by displacement damage in irradiated CCD detectors. The
development of the model took place in the highly challenging context of
{\gaia}, a European Space Agency mission scheduled for launch in 2012. {\gaia} will operate for 5 years in an orbit
around the second Lagrange point (L2) \citep{Gaia2001,Gaia2008} and will measure the parallaxes, proper motions, radial velocities, and astrophysical
parameters of over one billion stars. To do so, the satellite will constantly
scan the sky, observing the stars with two telescopes focussed on a single focal plane comprising 106 CCDs. The derived astrometric parameters are highly sensitive to the precise image shape, and hence to the effects of CTI.

{\gaia} will be subjected to the radiation environment at L2 which is
entirely dominated by protons emitted during solar flares. The energetic
protons collide with and displace atoms in the CCD silicon lattice, leading to
the creation of interstitial atom-vacancy pairs. The vacancies thus formed,
combine, by diffusion, with other vacancies or impurities (e.g. oxygen,
phosphorus, carbon atoms) present in the CCD as doping implants or
due to pollution during the fabrication process. The impurity-vacancy complexes
introduce energy levels in the semiconductor band gap that stochastically
trap and release the transferred signal carriers (electrons from the conduction band in n-type devices). The time-dependent capture and release probabilities
vary as a function of several factors; most importantly, the temperature, the
local charge density distribution in the vicinity of the trap, and trap parameters such as energy level and capture cross-section.

Based on the standard JPL model, the average proton dose received by the CCDs at the end of
the 5-year mission lifetime was originally predicted to be $4.14\times10^9$ protons cm$^{-2}$ (with 90\% confidence levels) for a launch in 2011 \citep{Fusero2007}. Current space weather forcasts predict that the next solar maximum may be considerably less severe than average, so that the dose may be rather lower. However the sensitivity of Gaia to radiation damage is such that this in no way reduces the need to calibrate the effects. In addition, the peak of the Solar activity is expected to occur in late 2013 which means that {\gaia} will receive
most of the total proton fluence early in the mission such that all data will be affected.

Based on experimental studies led by the industrial partners in the {\gaia}
project and independent analyses carried out within the {\gaia} science community, the
CTI resulting from radiation-induced traps is expected to affect
the mission performance by causing charge loss and image distortion.
Mitigating those effects has been recognized as critical to achieving the
mission requirements. Several aspects specific to {\gaia} contribute to the high
impact of radiation damage on the mission. The large focal plane is difficult to shield and the CCDs will be exposed to most of
the incoming particles. The required image location accuracy is extreme,
e.g., the end of mission parallax error is required to be better than $25$
micro-arcseconds for a star of magnitude $15$. The corresponding requirement on
the residual image location error per CCD transit is $0.01$~pixels. However the
image profile distortion induced by CTI has been measured to cause biases in the
image location measurement of up to $0.17$~pixels. {\gaia} will also study very
faint objects down to magnitude $20$ and at this signal level only a few electrons
comprise the PSF core and the effects of trapping are poorly understood.
{\gaia} will scan the sky by continuously spinning around an axis perpendicular
to the plane containing the telescope viewing directions
\citep[see][]{Gaia2001,Gaia2008}. In order to follow the resulting motion of the
stars across the focal plane and integrate the light during the transit, the
CCDs will be operated in Time-Delayed Integration mode (TDI mode). In this mode even
fairly bright objects will remain faint for a part of their transit. Likewise,
the sky background will form a gradient in the CCD parallel direction and, due
to the relatively short integration time, {\gaia} will not benefit from the potential trap filling effects of
a bright sky background.

These special operating conditions (high radiation dose, low signal level, low sky background and extremely high accuracy image location) demand a very high level of detail in the simulation of radiation damage effects, and preclude the use of models that assume instantaneous trapping within a certain volume that varies with the signal level \citep[e.g.,][]{massey2010, rhodes2010}. The Monte Carlo model presented here was developed to provide the required level of simulation detail. Our model thus simulates charge transfer at the electrode level and simulates the signal carrier trapping thanks to a new approach to the representation of both the charge density distribution and the capture and release probabilities. These simulations are used within the {\gaia} project to study the effect of CTI on measurements, to generate simulated data with which to verify the future radiation damage mitigation algorithms and to obtain a better understanding of CTI itself.
\newpage In the following sections we describe the relevant details of our Monte Carlo model and show that it can reproduce the experimental data obtained from
irradiated CCDs operated in TDI mode.

\section{Model description} \label{sect:model}

\begin{figure*}
  \includegraphics[width=1\textwidth]{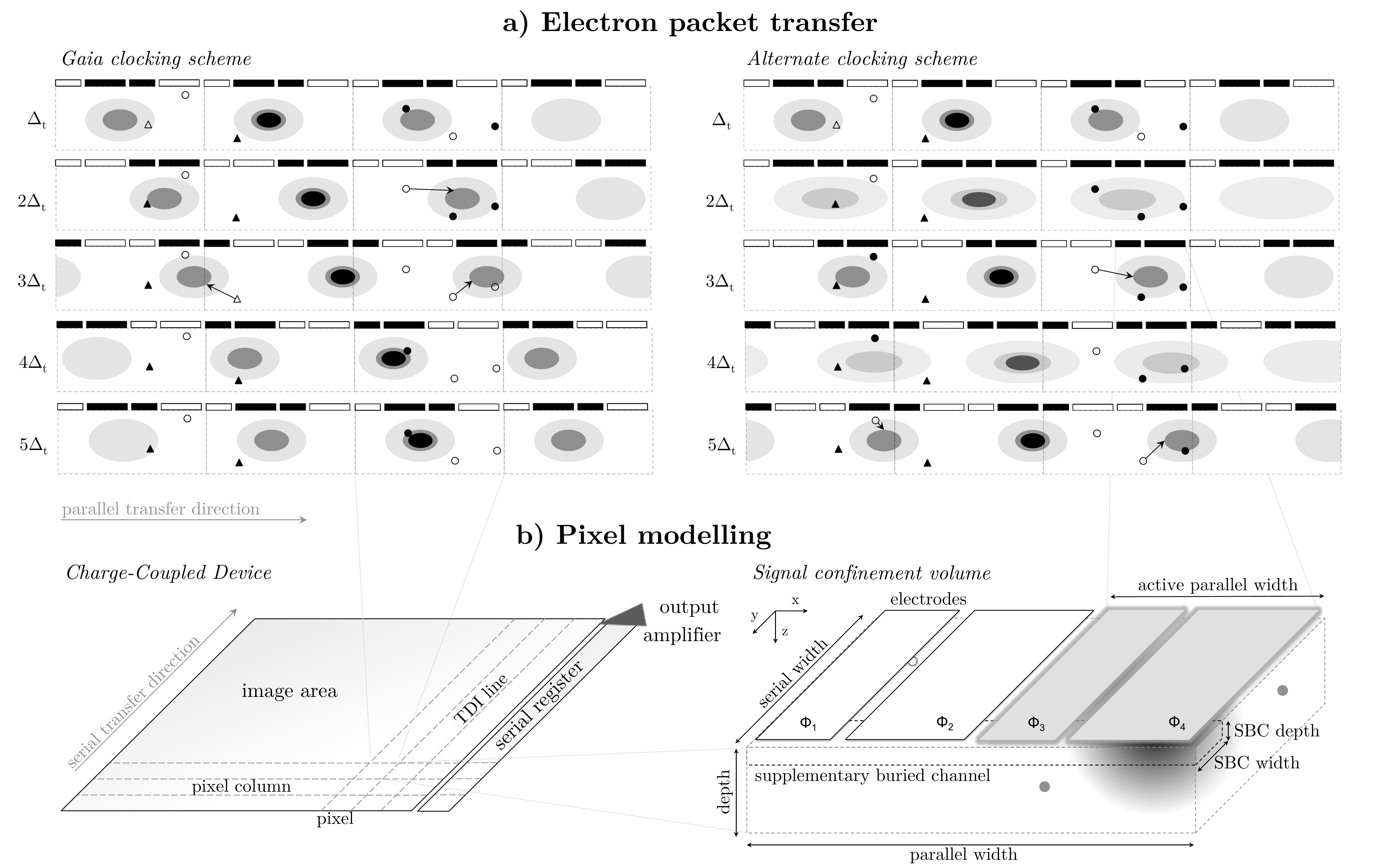}
  \caption{(a) depicts the transfer of several charge packets from pixel to
  pixel in a four phase device. A high voltage is applied on the black coloured electrodes (later referred to as `biased high')
  while no voltage is applied on the empty ones (at rest). The ellipsoidal shapes represent the charge
  density distribution under the electrodes. The packets interact with the traps
  of different trap species (circle and triangle) during the dwell time {\dwelltime}.
  The filled traps (filled black symbols) can release their charge (arrows) at
  any time. To illustrate the importance of an electrode level simulation, two
  different clocking schemes are shown; note the charge density distribution
  successively stretching and contracting in the alternate case compared to the
  {\gaia} case. \newline(b) illustrates the signal confinement volume, a key point of
  our modelling approach, and puts it in the context of the CCD as well as the
  supplementary buried channel. The CCD pixel rows are referred to here as TDI
  lines.\label{fig:schematic}}
\end{figure*}

As described by \cite{janesick2001}, a CCD needs to perform four fundamental
tasks to generate an image: charge generation, charge collection, charge
transfer, and charge measurement. The primary goal of our model is to simulate
the effects of charge traps, induced by displacement damage in the CCD. These traps affect principally the third fundamental task of a
CCD, the charge transfer, by stochastically capturing, and releasing charges during their transfer from one set of electrodes to another. The
charge collection process can also be affected if a trap present in the CCD
field free or depleted region captures a freshly photo-generated charge drifting
towards the CCD buried channel. We chose not to take into account this secondary
aspect and focus on the signal charge transfer only.  Thus our model
simulates exclusively the transfer of charges present in the signal confinement
region under each electrode in the image section and the serial register (Fig.\
\ref{fig:schematic}a). The signal confinement region is simulated as a box
(Fig.\ \ref{fig:schematic}b) of which the dimensions are defined by the
manufacturing characteristics of the CCD, i.e.\ the width of the electrodes biased high in the transfer direction,
the width of a pixel perpendicular to the transfer direction (serial direction) and in
depth by the depletion of an electrode biased high with no
electrons underneath. Fringing fields present at the edges of the signal
confinement region reduce the actual volume \citep{seabroke2008}.
To avoid any arbitrary assumption on the induced volume reduction, the
dimensions of the signal confinement volume remain set to the manufacturing
characteristics, while, as we shall see later in Sections \ref{sect:density}
and \ref{sect:fcl}, the actual distribution of the charge density within this
volume is constrained by experimental measurements. This compensates to some
extent our ignoring the fringing fields.

{\gaia} CCDs \citep{short2005} are custom made by e2v technologies and referenced as
CCD91-72. They are back-illuminated, full frame devices and incorporate a number
of specific features such as a charge injection structure and 12 TDI gates to
integrate bright stars over a shorter distance and avoid saturation.
Each pixel also contains a supplementary buried channel (SBC), and an
anti-blooming drain. In Table \ref{tab:ccdParam} important parameters of the
{\gaia} astrometric CCDs are summarized. Our model is capable of simulating the
effect of charge injection and also takes the SBC into account. The TDI gates and
anti-blooming drain are not explicitly modelled.

\begin{table}
\centering
 \begin{tabular}{|l|l|}
   \hline
   \textbf{Parameter} & \textbf{Value} \\
   \hline\hline
   \multicolumn{2}{|l|}{\textbf{General}}\\
   \hline
   Number of pixels (parallel $\times$ serial) & $4500\times1966$ \\
   Number of light sensitive pixels  & $4494\times1966$  \\
   Pixel size  (parallel $\times$ serial)  & $10\times30$ $\mathrm{\mu m^2}$\\
   Operational temperature & $163 \pm 3$ K\\
   \hline
   \multicolumn{2}{|l|}{\textbf{Image section}} \\ 
   \hline
   Number of phases  & 4 \\
   Transfer period  &  $982.8$ $\mu$s \\
   Pixel FWC  & $190\,000$ \electron\\
   SBC FWC$^{\star}$  & $\sim1300$ \electron\\
   SBC size$^{\star}$ ($1^\mathrm{st}$ CCD half) &  $10\times3$ $\mathrm{\mu m^2}$ \\ 
   SBC size$^{\star}$ ($2^\mathrm{nd}$ CCD half) &  $10\times4$ $\mathrm{\mu m^2}$ \\ 
   \hline
   \multicolumn{2}{|l|}{\textbf{Serial register}} \\
   \hline
   Number of phases  & 2 \\
   Transfer period &  $0.5$ $\mu$s (avarage) \\
   Pixel FWC & $475\,000$ \electron\\
   \hline
 \end{tabular}
 \caption{The e2v CCD91-72  parameters. SBC stands for supplementary buried channel and FWC for full well capacity. \newline $\star$ nominal values; the simulated and measured values differ significantly.}
 \label{tab:ccdParam}
\end{table}

\subsection{Simulation process} \label{sect:process}

\begin{figure}
  \centering \includegraphics[width=0.49\textwidth]{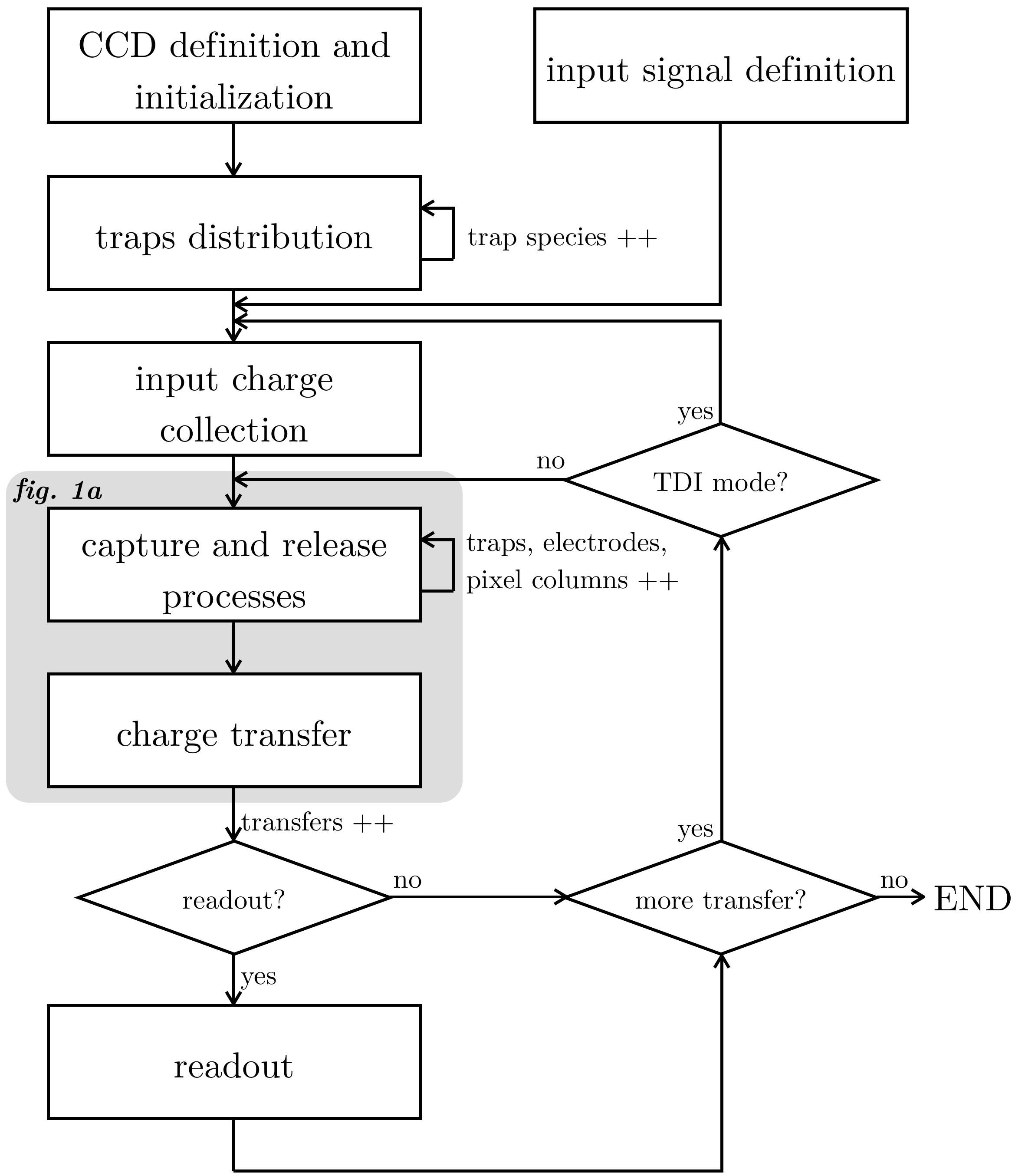}
  \caption{Top level diagram of the simulation process. \label{fig:diagram}}
\end{figure}

Figure \ref{fig:diagram} presents a simplified version of the whole simulation
process. The first step consists of defining an input signal and specifying the
CCD characteristics such as the number of pixels in the parallel and serial
directions, the number of electrodes per pixel, the clocking scheme, the
operating temperature and so on.

Prior to the actual trapping and transfer simulation, empty bulk traps are
randomly distributed across the CCD according to their specified concentration. Within each pixel, the traps are assigned a position in space by randomly
generating coordinates within the signal confinement volume. If necessary the
trap position can be kept fixed to repeat experiments with the exact same
simulated CCD. The traps can belong to different species defined by the
parameters described in Table \ref{tab:trapParam} according to the
Shockley-Read-Hall (SRH) formalism.

\begin{table}
  \centering
  \begin{tabular}{|l|l|}
    \hline
    \textbf{Parameter} & \textbf{Description} \\ \hline
    \hline
    \trapen\ (eV) & Energy level in the semiconductor gap \\
    \trapsigma\ (m$^{-2}$) & Capture cross section \\
    \trapdens\ (traps/pixel or m$^{-3}$) & Concentration \\
    \hline
  \end{tabular}
  \caption{The trap species parameters.\label{tab:trapParam}}
\end{table}

The charge collection step corresponds to the generation of charges under the
CCD electrodes. Photon detection from a light source
can be described as a Poisson process, thus we create photo electrons in the CCD
using a random generator with a Poisson distribution and a mean equal to the
expected number of collected photons within the integration time. As a
consequence only integer numbers of electrons are generated so that the actual
physical process is reliably simulated, including the variance in the number of
electrons generated. This is important in the {\gaia} context since a considerable
number of star images will contain only a few electrons per pixel due to the
faint nature of the observed sources and the operating mode of the CCDs. In TDI
mode, the light source motion is synchronized with the CCD charge transfer rate
so that the charge profile continues to build up as the image travels across the
CCD. The transfer period is equal to the integration time at each step, which
determines the number of photo-electrons generated. This implies that even for high
signal levels the number of charges transferred will be very low, at least
during the initial transfers. To simulate the background
illumination (sky-light and scattered light) the corresponding background count
rate is added to the expected number of collected photo-electrons before
invoking the Poisson random number generator. Electronic charge injections (CIs)
can also be simulated. In order to save simulation time any previous exposure
of the CCD to a continuous level of background illumination can be simulated analytically.
This is done by pre-computing (cf.\ Section \ref{sect:tol}) the corresponding trap occupancy level of each of
the trap species and filling the corresponding number of traps prior to the
transfer of the signal of interest.

The capture and release process is simulated as follows. From the trap
parameters, the density of charges at the trap position (cf.\ Section
\ref{sect:density}), and the interaction time, it is possible to calculate the
probability of capture or release (cf.\ Section \ref{sect:probabilities}), for
a specific trap relative to its state: empty or filled. If a trap is empty, the
capture probability $p$ is computed and a random number $R$ generated; if $R<p$, then the capture is triggered and a charge is removed from the charge packet. Correspondingly, for a full trap,
the release probability is computed and a random number is generated; if a charge
is released, it is added to the closest charge packet (cf.\ Fig.\
\ref{fig:schematic}a). This procedure is repeated for each trap in the CCD,
pixel column by pixel column. The interaction time (cf.\
Section \ref{sect:probabilities}) is defined by the amount of time a charge packet stays
under a given set of electrodes. It defines the temporal resolution of the
simulation and depends on the charge transfer period, the number of electrodes
and the clocking scheme. CCDs with two, three, or four phases can be simulated,
and any kind of clocking scheme applied. This facilitates, for instance, testing
the radiation hardness of different CCD configurations, taking into account the
specific measurements to be carried out.

During the charge transfer step, the CCD electrodes are biased high or set at rest
according to the predefined clocking scheme. The charge packets are then
redistributed under the next set of biased high electrodes. Trapping, transfer and
charge collection (in TDI mode) are repeated until
the last charge packet belonging to the input signal reaches the serial
register, at which point the simulation ends.

During read-out the signal charges are collected. It is also possible to
simulate the charge and release processes in the serial register by repeating
the same procedure as for the image section and making sure the illumination
history is respected, i.e.\ all pixels in a line are processed, ordered by distance from the output amplifier (Fig.\ \ref{fig:schematic}b).

\subsection{Effective charge capture and release probabilities} \label{sect:probabilities}

In the SRH formalism charge capture and release are described as decay
processes. One can derive the charge capture and release probabilities as
follows. First let us consider a number of filled traps \nfull. In an
infinitesimal time interval $dt$, the number of released charges is proportional
to \nfull. The proportionality constant is the release rate \emrate. The number
of filled traps as a function of time can then be derived:

\begin{equation}
  \begin{aligned}
    \frac{d\nfull}{dt} &= - \emrate\nfull\,, \\[3pt]
    \nfull(t) &= N_{\mathrm{full},0} e^{-\emrate t}\,,
  \end{aligned}
  \label{eq:variation}
\end{equation}

where $N_{\mathrm{full},0}$ is the number of filled traps at $t=0$. The fraction
of filled traps remaining after a time $t$ is then statistically equivalent to
the probability for any specific trap to remain filled after a time $t$:

\begin{equation}
  \frac{\nfull(t)}{N_{\mathrm{full},0}} = \pfull(t) = 1 - \prel(t) =
  e^{-\emrate t}\,,
\end{equation}

where {\prel} is the probability that a filled trap releases an electron within a
time interval $t$:
\begin{equation}
  \prel(t) = 1 - e^{-\emrate t}\,.
\end{equation}

The release rate constant is given by:
\begin{equation}
  \emrate = \frac{1}{\taurel} = X \chi \trapsigma v_\mathrm{th} n_\mathrm{c} e^{- \trapen / kT}\,,
\end{equation}

where {\taurel} is the release time constant, {\trapsigma} the capture
cross-section, {\trapen} the trap energy level in the semiconductor forbidden
gap, $X$ the entropy factor, $\chi$ the field enhancement factor, $T$ the CCD
operating temperature, $k$ the Boltzmann constant, and $v_{th}$ the electron
thermal velocity:

\begin{equation}
  v_\mathrm{th} = \sqrt{\frac{3kT}{m^*_\mathrm{e}}}\,.
\end{equation}
with $m^*_\mathrm{e} = 0.5\,m_\mathrm{e}$ the effective electron mass. The effective density of
states $n_\mathrm{c}$ in the conduction band is:
\begin{equation}
  n_c = 2\left(\frac{2\pi m^*_e k T}{h^2}\right)^{3/2}\,,
\end{equation}
where $h$ is the Planck constant.

Likewise one can derive the probability that an empty trap captures an electron
within a time interval $t$:
\begin{gather}
  \pcap(t) = 1 - e^{-\caprate t} \\[3pt]
\caprate = \frac{1}{\taucap} = \trapsigma v_\mathrm{th} \edens
  \label{eq:captureTime}
\end{gather}
where {\caprate} is the capture rate, {\taucap} the capture time constant, and
{\edens} the electron density at the trap location (cf.\ Section
\ref{sect:density}).

The characteristic interaction time between the traps and a charge
packet is the dwell time {\dwelltime}. It corresponds to the elapsed time between two
charge redistributions, during which {\edens} remains constant. The dwell time
is proportional to the transfer period and depends on the selected clocking
scheme. {\dwelltime} is greater in the CCD image area than in the serial register,
since a complete TDI line ($\sim2000$ pixels) must be read out during a pixel to
pixel transfer in the image area. For the {\gaia} CCDs, {\dwelltime} varies from
several tens of micro-seconds (serial register) up to a forth of milli-second (image
section). The release time constants for certain trap species can be as short as
several hundreds of nano-seconds like the A centre (oxygen-vacancy complex).
These `fast' traps can lead to multiple capture and release events during a
dwell time and play an important role in the CTI associated with the serial
register. As a consequence we shall follow \cite{LL:SAG-LL-022} and introduce
effective probabilities of charge capture and release that take into account the
possibility of multiple capture and release cycles within one time interval. We
consider the probability {\pfull} for a trap of unknown state to be filled after
a time interval $t$. $N$ is the total number of traps and {\nempty} the number
of empty traps:
\begin{equation}
  N = \nfull + \nempty
\end{equation}
We follow the same derivation as in (eq. \ref{eq:variation}) but now accounting
for the empty traps
\begin{equation}
  \begin{aligned}
    \frac{d\nfull}{dt} &=\caprate\nempty - \emrate\nfull\,, \\
    \frac{1}{N} \frac{d\nfull}{dt} & = \frac{d\pfull}{dt}\,, \\
    & = \caprate\pempty - \emrate\pfull\,, \\
    & = \caprate (1-\pfull) - \emrate\pfull \,.
  \end{aligned}
\end{equation}
In these equations {\pempty} is the probability for a trap of unknown state to
be empty after the time interval $t$. One can then derive the following general
solution, assuming that {\caprate} and {\emrate} are constants:
\begin{equation}
  \pfull(t) = \frac{\caprate}{\emrate+\caprate} + C \exp\left[ -\left(\emrate
  +\caprate\right)t \right]\,.
  \label{eq:sol1}
\end{equation}
It is now possible to derive the effective capture and release probabilities,
{\pcap} and \prel. If the trap is empty at $t=0$ then $\pfull(0)=0$ and
$C=-\caprate/(\emrate+\caprate)$, thus:
\begin{equation}
  \pcap \equiv \pfull(t) = \frac{\caprate}{\emrate+\caprate}
  \left(1-\exp\left[-\left(\emrate + \caprate\right)t \right] \right)\,.
  \label{eq:captureproba}
\end{equation}
Similarly, if the trap is filled at $t=0$ then $\pfull(0)=1$ and
$C=+\caprate/(\emrate+\caprate)$. Hence:
\begin{equation}
  \begin{aligned}
    1 - \prel \equiv \pfull(t) = \frac{\caprate + \emrate \exp\left[
    -\left(\emrate + \caprate\right)t \right]}{\emrate+\caprate}\,, \\[3pt]
    \prel = \frac{\emrate}{\emrate+\caprate} \left(1-\exp\left[ -\left(\emrate +
    \caprate \right)t \right] \right)\,.
  \end{aligned}
\end{equation}

\subsection{Charge density distribution modelling} \label{sect:density}

The CTI effects model we describe in this paper is a density driven model to
be contrasted with the more commonly used volume driven models. The volume
driven models assume instantaneous trapping within a certain volume that
varies with the signal level. A density driven model necessitates the
computation of the capture and release probabilities for each trap --- taking
the charge density in the trap vicinity into account --- regardless of its
location (no trap is a priori ignored). The density driven model thus requires
the evaluation of the charge density distribution as a function of the signal
level and location within the pixel signal confinement region. The confinement
region is defined by the electrodes biased high. As can be seen from the work
of \cite{hardy1998} and more recently \cite{massey2010} and \cite{rhodes2010},
the volume driven approach is fairly successful in explaining experimental
data, in particular {\hst} data. However the {\gaia} operating conditions
differ significantly from the {\hst} ones. The {\gaia} CCDs will be operated
in TDI mode. In this mode the exposure time equals the charge transfer period.
Thus the charge-trap interaction time is significantly decreased and
instantaneous trapping cannot be assumed anymore. Moreover, as already stated
in the introduction, {\gaia} will deal with very low levels of background and
source signal. Trapping in these particular conditions was investigated for
the first time in studies related to the {\gaia} mission. The density driven
approach proved to be necessary to explain and reproduce experimental results
\citep{short2007,seabroke2008} which show that the CTI effects are modified by
an extremely small level of background light (of the order of a few photons
per second). The volume occupied by these very few electrons should be
negligible and thus prevent any trapping from occurring according to
volume-driven models. However the {\gaia} experimental studies showed that
very few electrons are capable of filling a significant amount of traps. This
can be explained by a density driven approach. Indeed, as the background light
constantly illuminates the CCD, the long effective charge-trap interaction
time for these very few electrons compensates for the very small charge
density in the vicinity of each trap and trapping then becomes likely to
occur. Our simulation method is also particularly convenient for accurately
simulating CTI effects over a wide range of signal levels and for CCDs in
which extra doping implants (such as a supplementary buried channel) modify
the pixel potential and induce non-linearities between the CTI effects and the
signal intensity (cf.\ Section \ref{sect:validation}).

The charge density distribution within the signal confinement volume, which
varies strongly with the CCD architecture, is thus a key parameter in CTI
modelling. Although the distribution cannot be directly measured, in principle
it can be accurately determined for a specific CCD architecture and signal level
by solving simultaneously the Poisson equation and the charge continuity
equation \citep[e.g.][]{seabroke2009} in order to find a consistent electrode potential
and charge carrier distribution. This requires a detailed knowledge of the CCD
implant characteristics (i.e., the nature and concentration of the dopants).
This information is often commercially sensitive and in order to keep our model
flexible regarding its application to other cases than {\gaia}, we use an analytical
description of the charge density distribution which is roughly consistent with
the modelling results by \cite{seabroke2009}. This analytical description
consists of a normalized Gaussian function in the three space directions of
which the complexity increases with the number of the CCD potential
characteristics included. The distribution parameters are listed in Table
\ref{tab:densityParameters}. The density is defined as:
\begin{equation}
  \edens(x,y,z) = S\times\rho(x,y,z) = S\times \rho(\vect{x})\,,
\end{equation}
where
\begin{equation}
  \rho(\vect{x}) =
  \frac{\exp\left[-\frac{1}{2}\left(\vect{x}-\vect{x}_0\right)^T {\bf C}
  ^{-1}\left(\vect{x}-\vect{x}_0\right)\right]}{\left(\sqrt{2\pi}\right)^3
  \left|{\bf C}\right|^{1/2}}\,,
\end{equation}
with
\begin{equation}
  {\bf C} = \begin{pmatrix}
    \sigma^2_{x} & 0 & 0 \\
    0 & \sigma^2_{y}  & 0 \\
    0 & 0 & \sigma^2_{z} \\
  \end{pmatrix}\,.
\end{equation}
Hence
\begin{equation}
  \rho(\vect{x}) =  \frac{
  \exp\left[-\frac{1}{2}\left(\left(\frac{x-x_0}{\sigma_x}\right)^2 +
  \left(\frac{y-y_0}{\sigma_y}\right)^2+\left(\frac{z-z_0}{\sigma_z}\right)^2\right)\right]}{
  \left(\sqrt{2\pi}\right)^3\sigma_x\sigma_y\sigma_z}\,.
  \label{eq:simpleGaussian}
\end{equation}
\begin{table}
  \centering 
  \begin{tabular}{|l|l|} 
    \hline
    \textbf{Parameter} & \textbf{Description} \\
    \hline\hline
    $S$ (\electron) & number of \electron in the signal confinement volume \\
    \nsat\ (m$^{-3}$) & density saturation level\\
    \hline
    \multicolumn{2}{|l|}{\textbf{Signal confinement volume} } \\
    \hline
    $x_\mathrm{max}$ (m) & parallel width\\
    $y_\mathrm{max}$ (m) & serial width\\
    $z_\mathrm{max}$ (m) & depth \\
    $V$ ($\mu$m$^3$) & signal confinement volume $x_\mathrm{max}\times
    y_\mathrm{max} \times z_\mathrm{max} $\\
    \hline
    \multicolumn{2}{|l|}{\textbf{Buried channel regime} } \\
    \hline
    $\sigma_x$ (m) & parallel distribution standard width\\
    $\sigma_y$ (m) & serial distribution standard width \\ 
    $\sigma_z$ (m) & depth distribution standard width \\
    $x_0$ (m) & distribution centre parallel coordinate\\
    $y_0$ (m) & distribution centre serial coordinate\\
    $z_0$ (m) & distribution centre depth coordinate\\
    \ssat (\electron)& Signal level at which the buried channel saturates $\equiv$  FWC\\
    \hline
    \multicolumn{2}{|l|}{\textbf{Supplementary buried channel regime}} \\
    \hline
    $\sigma_{y,\mathrm{SBC}}$ (m) & parallel distribution standard width\\
    $\sigma_{z,\mathrm{SBC}}$ (m) & depth distribution standard width\\
    $y_{0,\mathrm{SBC}}$ (m) & distribution centre serial coordinate\\
    $z_{0,\mathrm{SBC}}$ (m) & distribution centre depth coordinate\\
    $S_\mathrm{SBC}$ (\electron) & Signal level at which the SBC saturates $\equiv$  FWC$_\mathrm{SBC}$ \\
    \hline
  \end{tabular} 
  \caption{Charge density distribution parameters} 
  \label{tab:densityParameters} 
\end{table}
\begin{figure}
  \includegraphics[width=0.49\textwidth]{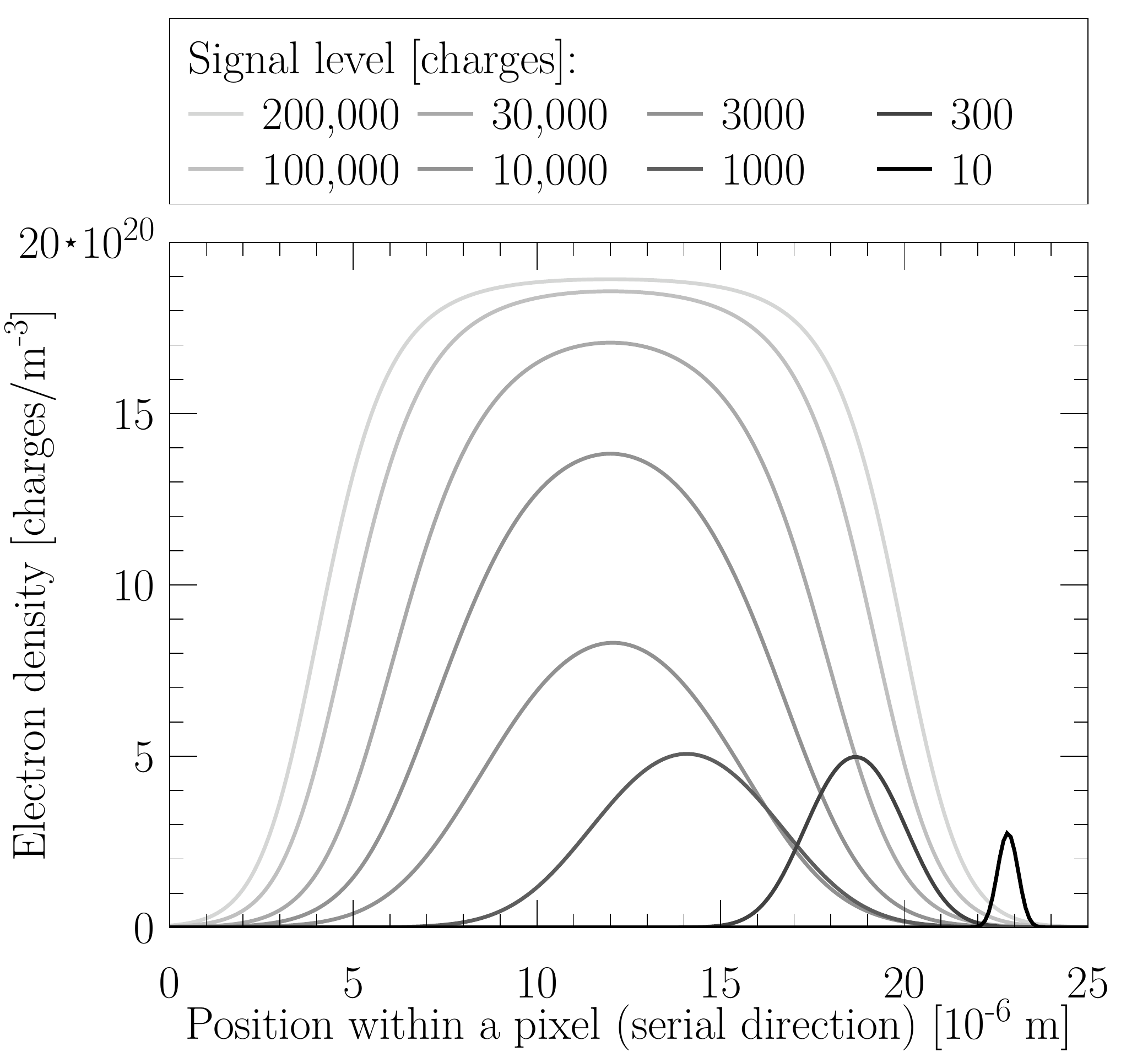}
  \caption{Example of a charge density distribution as it is simulated by the presented analytical model: the charge density at the center of the charge cloud (in the parallel direction and in depth) is plotted along the CCD serial direction for different signal levels. The simulation includes an off-centred SBC and a saturation occurring at high signal level. \label{fig:density}}
\end{figure}

The supplementary buried channel (SBC) corresponds to an additional doping
implant, which generates a deeper potential well and narrows the charge
distribution so that electrons are transferred through a smaller volume of silicon and encounter fewer traps. In the {\gaia} CCDs, the SBC
potential well collapses at signal levels of a few thousand electrons. As a consequence, the SBC only improves CTI in the small signal regime. To implement the SBC, we simply introduce a second, low signal regime with
a different charge density distribution and ensure a smooth transition between
them so that (eq. \ref{eq:simpleGaussian}) becomes:
\begin{equation}
  \rho(\vect{x}) =  \frac{
  \exp\left[-\frac{1}{2}\left(\left(\frac{x-x_0}{\sigma_x}\right)^2+ 
  \left(\frac{y-y_0^\star}{\sigma_y^\star}\right)^2+
  \left(\frac{z-z_0^\star}{\sigma_z^\star}\right)^2\right)\right]}{
  \left(\sqrt{2\pi}\right)^3\sigma_x\sigma_y^\star \sigma_z^\star}
  \label{eq:sbc}
\end{equation}
where the parameters indicated with a $\star$ now vary as a function of signal
level:
\begin{equation}
  P^\star = P \left(1 - e^{-\frac{S}{S_\mathrm{SBC}}}\right) +
  P_\mathrm{SBC} e^{-\frac{S}{S_\mathrm{SBC}}}\,,
\end{equation}
where $P$ refers to the parameter value in the buried channel and
$P_\mathrm{SBC}$ to the corresponding value in the SBC.

At high signal levels, saturation effects limit the linear growth of the charge
density; while more charges can still be added, the maximum density cannot be
overcome and the distribution expands in the three spatial directions.
Saturation occurs when the charge density becomes larger than the doping
concentration. Taking into account the presence of the anti-blooming drain
we consider that locally, the charge density cannot exceed
$\nsat = \ssat/V_e$, with $V_e= (2\pi)^{3/2} \sigma_x\sigma_y\sigma_z$.
Following the method proposed by  \cite{LL:SAG-LL-022}, the model for
the saturation process is:
\begin{equation}
  \edens(\vect{x}) = \frac{\nsat S^\prime\rho(\vect{x})}
  {\nsat + S^\prime\rho(\vect{x})}
\end{equation}
where $S^\prime$ has to be adjusted to give the correct total
charge $S$:
\begin{equation}
  \begin{aligned}
    S = \iiint \edens(x,y,z)\,dx\,dy\,dz\,,\\[3pt]
    S = \sqrt{\frac{2}{\pi}} \int^\infty_0 \frac{ u r^2}{u + \exp\left( r^2/2
    \right)}\, dr\,,
  \end{aligned}
\end{equation}
with $u =S^{\prime}/\ssat$ and $\ssat=\nsat\times(2\pi)^{3/2} \sigma_x\sigma_y
\sigma_z$. In the linear growth case:
\begin{equation}
  u \ll 1 \quad\mathrm{and}\quad S\simeq \ssat u\,.
\end{equation}
In the saturation case:
\begin{equation}
  u \gg 1 \quad\mathrm{and}\quad S\simeq \ssat \left(4/3 \sqrt\pi
  \right)\left(\ln u\right)^{3/2}\,.
\end{equation}
The adjustment of $S^\prime$ is then analytically constructed to provide the
right behaviour in the linear and saturation cases and a reasonable
approximation for $u \approx 1$, i.e.\ the transition region:
\begin{equation}
  S^{\prime} = S \left( 1 + \left( S/\ssat\right)^{0.8}\right)^{-1.25} e^{\left(
  \frac{3\sqrt{\pi}}{4} S/\ssat \right)^{2/3}}\,.
\end{equation}

Figure \ref{fig:density} shows an example of the evolution of the charge density distribution as a function of the signal level as it is simulated by the presented model. It illustrates the particular features that the model aims to reproduce. At low signal level the charge density distribution shows a narrow profile in accordance with the expected SBC effect. While the signal level increases, the density profile widens and shifts to the centre of the pixel. It corresponds to the transition from the SBC regime to the BC regime in the case of a SBC implant not located at the centre of the pixel as depicted in Fig.\ \ref{fig:schematic}b (which corresponds to the \gaia\,case). And finally one can clearly observe the saturation occurring for the large signal levels.

\subsection{CCD illumination history} \label{sect:tol}

The CCD illumination history determines the CCD state (i.e.\ the trap occupancy
level) prior to a star transit. Every star brighter than
magnitude $20$ will be observed $\sim80$ times by {\gaia} (on average). During each observation of a given star, the satellite
will be scanning the same part of the sky but from a different direction i.e. with a different orientation of the focal
plane. Hence the occupancy of the traps in the CCD will be different prior to each transit of a
given star. As a consequence the CTI effects (image location estimation
bias and charge loss) are likely to be different from one observation of a
particular star to the next (even if the radiation damage to the CCDs were unchanged). Apart from the trapping process and the electron density
distribution, the CCD illumination history is therefore another key element of
the CTI effects modelling. The illumination history is determined by discrete events
such as star transits and charge injections, which are directly reproduced
during the simulation. Our simulations also account for a continuous optical background comprising light from unresolved
stars and scattered light within the spacecraft. The light from the sky background constantly illuminates the {\gaia} CCDs at
the level of $\sim42$ \electron\ s$^{-1}$ arcsec$^{-2}$ (taking the
contributions of both telescopes into account), i.e.\ each CCD pixel will on
average receive $\sim 4\times10^{-4}$ \electron\ per pixel transfer step. In TDI mode the
charges generated by the sky background are not only integrated but also
continuously transferred from one pixel to another, to form a signal gradient
along the parallel direction, the last light-sensitive pixels in the CCD
effectively receiving about $2$ \electron.

\cite{swb2009b} showed that such low levels of constant illumination noticeably
modify the trap occupancy level. Indeed even if the local electron density
distribution remains approximately unchanged, the interaction time of these
background electrons with the traps is effectively much greater than the one of
a usual transiting source. To simulate the effect on the trap occupancy level,
one can run the model for a certain amount of time with the background light as
the only input signal. The star signal of interest is then inserted once an
equilibrium trap occupancy level has been attained. However, our model is
computationally very intensive and using it to simulate several minutes of CCD
operation would take too much time. For this reason prior to each simulation, we determine the state of each trap using a random generator and considering the probability $\pinf$, the probability for a particular trap to be filled by an electron from the background light. $\pinf$ is computed from eq.\ \ref{eq:captureproba} assuming an infinite time of interaction:
\begin{equation}
\pinf = \pfull(t = \infty) = \frac{\caprate}{\emrate+\caprate}.
\end{equation}
In the simulations, the background contribution is also taken into
account during the charge collection step by adding the background count rate to
the expected number of collected photo-electrons before invoking the Poisson
random number generator (cf.\ Section \ref{sect:process}). Note that the dark current can be taken into account in the same way. However in the {\gaia} case, the contribution of the dark current to the global background at operational temperature is not significant ( $\sim 10^{-7}$ \electron per pixel transfer step), we thus ignore it in the following.

\section{Radiation tests} \label{sect:tests}
We describe the verification of our model against experimental data.
These data were obtained by the industrial partners in the {\gaia} project. 

Several experimental studies were carried out on irradiated CCDs in order to
evaluate the impact of CTI on {\gaia}'s scientific requirements, to define the
optimal operating temperature, to prepare the CCD calibration activities, and to
elaborate a radiation damage mitigation strategy. Sira electro-optics acquired
the first sets of CCD radiation test data \citep{hopkinson2005}, focusing on the
determination of trap parameters and CCD characterization. Later, Surrey
Satellite Technology Limited (SSTL, formerly Sira) investigated the potential difference between
a CCD irradiated at room temperature and a CCD irradiated at $163\mathrm{K}$ and
kept at that temperature. SSTL concluded that the results obtained for CCDs irradiated at room temperature should be adequate for {\gaia} performance
predictions within the usual experimental uncertainties \citep{surrey2008}.

Up to now the prime contractor for {\gaia}, EADS-Astrium, has performed three test campaigns on {\gaia} CCDs irradiated at room temperature with a
radiation dose of $4\times10^9$ protons cm$^{-2}$ ($10$ MeV equivalent), and a fourth one is on-going. The
experimental setup includes a translation stage which enables it to reproduce the
star motion and hence to operate the CCD in TDI mode. The CCD was cooled and operated at
constant temperature throughout the test campaigns. During the first campaign,
Astrium concentrated on determining CTI effects on {\gaia}'s
astrometric measurements: the charge loss and the image profile distortion
leading to a biased evaluation of the image location. The experiments were
carried out as a function of stellar brightness (i.e.\ signal level) and
background light level. The purpose of this first campaign was also to evaluate
the viability of an artificial diffuse optical background source as a CTI mitigation device. The
second campaign (RC2) allowed an alternative mitigation tool, Charge Injection (CI), to be thoroughly
studied. Each of {\gaia}'s CCDs contains a row of diodes and gate
electrodes before the first pixel electrodes in the image section. A row of
charges can thus be injected and transferred to fill the traps prior to a star
transit. The CI level or the quantity of injected electrons is defined by the
difference in voltage applied to the pixel and the gate electrodes. To study the
influence of the CI parameters on the CTI effects, RC2 data has been acquired
for different CI levels, durations (number of CI rows at a time), and delays
(elapsed time between the CI and the first star transit), at different
temperatures. The third campaign used a realistic sky-like illumination pattern
to simulate the star transits instead of the uniform illumination grid from RC2.

Not only astrometric tests were performed during these campaigns but photometric and spectrometric issues were also addressed. The CCDs used by the red photometer and the radial velocity spectrometer (RVS) instruments on board {\gaia} differ slightly from the astrometric ones. They are based on the same architecture but are thicker devices and use a modified anti-reflection coating to enhance their quantum efficiency in the red wavelength band. A set of preliminary tests was performed during RC2 on such a device irradiated with a lower radiation dose ($2\times10^9$ protons cm$^{-2}$), followed by a more detailed study during the radiation campaign 3 (RC3). This study included a mask mimicking realistic G2V stellar spectra and a very tight control over the background light. As a result, the red-enhanced CCD was established to be more sensitive to radiation damage (most likely due to a greater depletion depth) and the expected shift and distortion of the spectral features induced by CTI were characterized as a function of stellar brightness and level of background light.

The raw data acquired during these campaigns were made available to the {\gaia} Data Processing and Analysis Consortium in order to
be re-analyzed and to support the CTI modelling efforts.

To compare our model results with the experimental data, we relied mostly on the
RC2 data \citep{astrium2008,swb2009}. We now describe in more detail RC2
experiments. The CCD irradiation scheme of the image section consists of three areas of similar size. The first area is non-irradiated, the second is
irradiated at the level of $4\times10^9$ protons cm$^{-2}$ ($10$ MeV equivalent)
and the third, irradiated at a higher dose, was not used during testing. The
serial register is non-irradiated. The reference and CTI affected data are
obtained by translating the light source and a mask over respectively the first
and second area of the CCD. The mask contains $50\times22$
(parallel$\,\times\,$serial) holes, and the projected stellar images are
separated by 50 pixels in the parallel direction and 20 pixels in the serial
direction. Each stellar image is binned along its serial dimension which
necessitates some assumptions in the input signal modelling (detailed in Section
\ref{sect:validation}). For each test at least $5$ consecutive scans with
identical configuration are performed. The mean time interval between the end of
a scan and the beginning of a new one is $29.7$ s. The first scan has a
different illumination history from the others and is usually excluded from the
analysis. We used two different astrometric tests out of the three that were
performed during RC2. In the first test the CI delay and duration are kept fixed
while the CI levels vary from $\sim4000$ to $\sim115\,000$ \electron. Each injection level was tested at different
temperatures (from $163$ K to $198$ K) and for different illumination levels
(corresponding to stellar magnitudes $13.3$ and $15$). In the second test the
delay between the CI and the first star transit is varied from $30$ to at most
$120\,000$ pixels (i.e.\ from $\sim29$ ms to $\sim118$ s) while the CI level is
fixed to $20\,000$ {\electron} and the CI duration to $20$ pixels. The test
temperature was kept close to $163$ K. The whole sequence of tests was repeated for different illumination
levels (star magnitudes $13.63$, $15.29$, $16.96$, $18.65$, $20.25$). In the
following section, we explain how we validated our MC model against these
experimental tests.

\section{Model verification and comparison to experimental data} \label{sect:validation}
\begin{figure*}
  \includegraphics[width=0.49\textwidth]{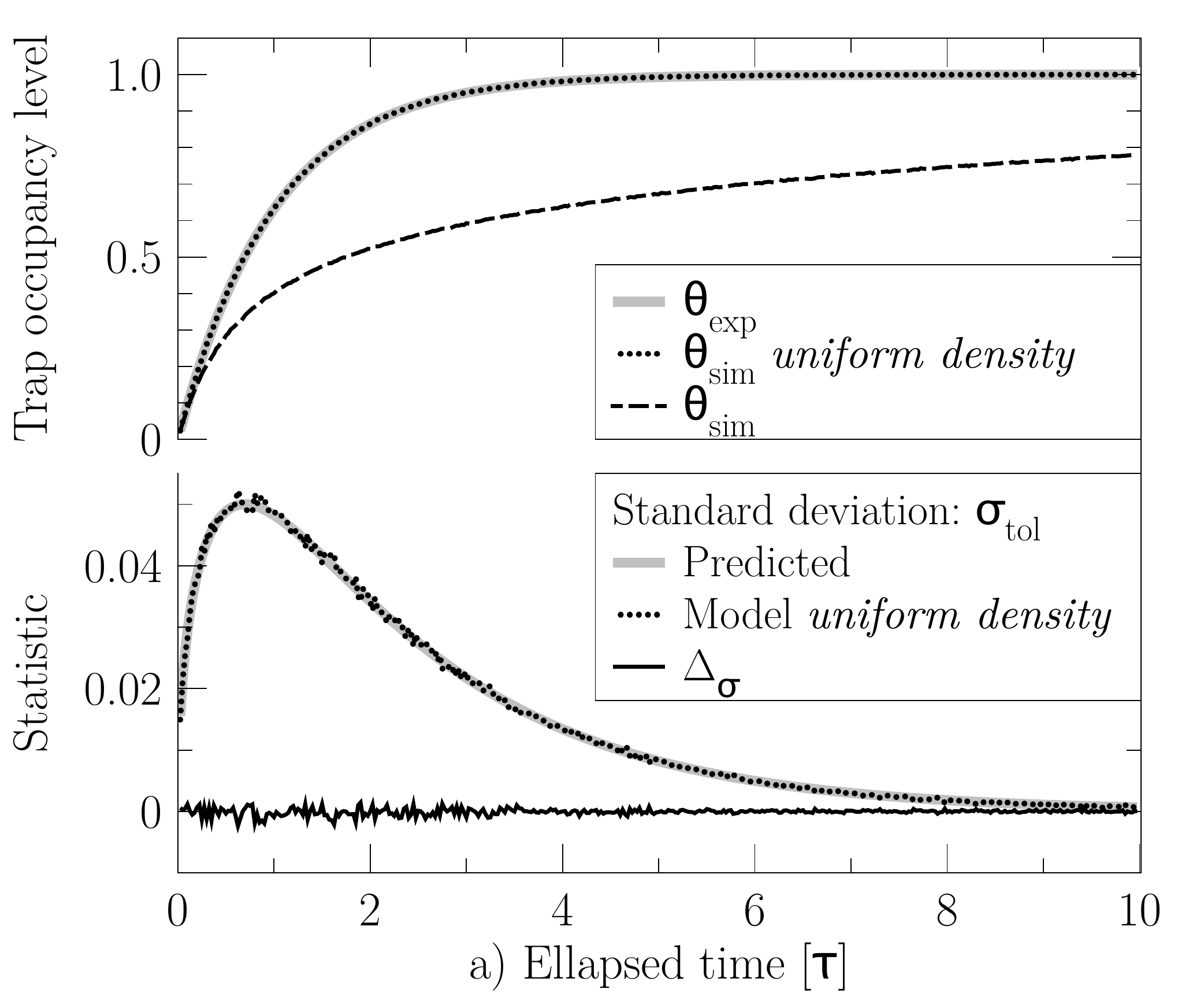}
  \includegraphics[width=0.49\textwidth]{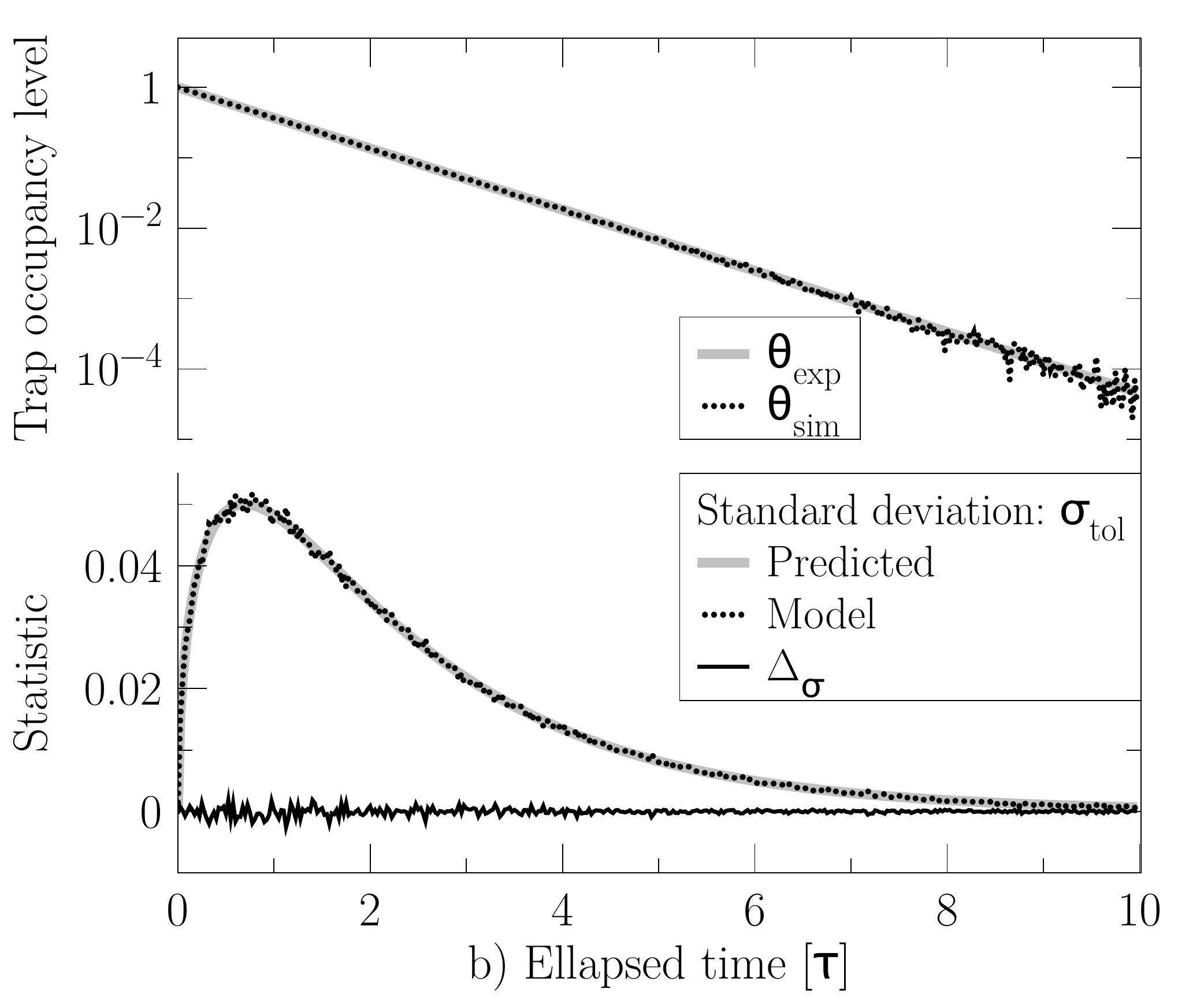}
  \caption{Tests of the capture (a) and release (b) processes. Top panels: comparison
between the expected trap occupancy level (grey) and the mean trap occupancy
level for 1000 Monte Carlo realizations for a CCD containing 100 traps
(black). For the capture test (a), the dotted line depicts the results
obtained for a uniform electron density distribution and the dashed line for a
3D Gaussian distribution including a SBC (eq.\ \ref{eq:sbc}). Bottom: the
expected (grey line) and actual standard deviation (dotted black line,
measured from the Monte Carlo experiments) for the trap occupancy levels, and
the difference between them (black line). \label{fig:basic}}
\end{figure*}
To validate the model we proceed in two steps. First the unit
blocks of the model, the capture and release processes, are tested individually
by comparing the results of
Monte Carlo simulations to statistical analytical predictions derived from the
equations in Section \ref{sect:probabilities}.
The second step consists of a direct comparison between the model and a subset
of the RC2 data. The variety of tests performed under different experimental
conditions during RC2 allows us to separately verify the different features of
our model. We first address the capability of our electron density distribution
model to reproduce the CTI effects over a large range of signal levels and in
particular to mimic the effect of the CCD supplementary buried channel.
Subsequently we investigate how well the model reproduces individual measurements
acquired in TDI mode.

\subsection{Model comparison to analytical prediction} \label{sect:verif1}

During each transfer step of a simulation, it is possible to monitor the trap
occupancy level {\thetasim} (or fraction of filled traps in the CCD):
\begin{equation}
  \thetasim = \frac{\nfull}{N}\,,
\end{equation}
with {\nfull}, the number of filled traps and $N$, the total number of
traps in the CCD.

For a simple experimental configuration, one can analytically predict the
evolution of the trap occupancy level {\thetaexp} as a function of time and
compare it with {\thetasim} to verify the reliability of the basic steps in the
Monte Carlo simulation. The capture module in the simulation is tested by
simulating a CCD under a constant illumination, which contains a unique trap
species with a capture time constant {\taucap} and an infinite release time
constant so that no electron release can occur. We also make all traps in the
CCD empty at the beginning of the simulation, i.e.\ $\thetasim(0)=0$. Under
these conditions, the expected trap occupancy level {\thetaexp} is equivalent to
the capture probability: 
\begin{equation}
  \thetaexp(t) = \pfull = \pcap = 1 - e^{-\frac{t}{\taucap}}\,.
\end{equation}
To compute {\taucap} analytically (eq. \ref{eq:captureTime}), we assume a
uniform electron density distribution, which we also implemented in our model
for the purpose of this test.
Similarly, to test the release module in the simulation we simulate a CCD in complete
darkness, which contains a unique trap species with a release time constant, 
{\taurel} and an infinite capture time constant. All the traps are artificially
filled at the beginning of the simulation so that $\thetasim(0)=1$. Under
these conditions, one can write: 
\begin{equation}
  \thetaexp(t) = \pfull = 1 - \prel =  e^{-\frac{t}{\taurel}}
\end{equation}

The number of traps in a CCD is finite. Thus, {\thetasim} and {\thetaexp} are
necessarily different due to the discrete nature of the capture and release
process (no fractions of electrons can be captured or released). Yet we want to
assess how accurately our model reproduces the expected value in these
particular conditions. After a time $t$ each trap in a CCD can only be empty or
filled, with the probability of occupancy given by {\pfull}. Hence the simple
simulations of trapping or release considered here constitute a Bernoulli trial where
one counts the number of times $n$ that a trap is filled or emptied after time
$t$. The value of {\thetasim} is then given by $n/N$ and will follow a binomial
distribution with the following expectation value and variance:
\begin{equation}
  \begin{aligned}
    E(\thetasim) = \pfull = \thetaexp\,, \\[3pt]
    \mathrm{Var}(\thetasim)= \frac{1}{N} \pfull (1 - \pfull) =
    \frac{1}{N} \thetaexp (1 - \thetaexp)\,.
  \end{aligned}
  \label{eq:sd}
\end{equation} 
We repeat the Monte Carlo simulation $N_\mathrm{run}$ times in order to compute
the mean trap occupancy level $\langle\thetasim\rangle$ and its variance
$\mathrm{Var}(\thetasim)$ and to verify whether these values are the same as
the expected ones given in (eq. \ref{eq:sd}).  As can be seen from Fig.
\ref{fig:basic}, for the capture and release tests the model reproduces the
expected results.
It is also interesting to note that even in this simplistic experimental
configuration there is a clear difference (Fig. \ref{fig:basic}a) between the capture profile obtained
for a uniform electron density distribution and a more realistic distribution as
described in Section \ref{sect:density}.

\subsection{Model comparison to experimental data}  \label{sect:verif2}

\subsubsection{Fractional charge loss} \label{sect:fcl}

The charge loss induced by CTI is directly connected to the trap capture
process, which is particularly sensitive to the electron density distribution.
As a consequence, reproducing charge loss measurements over a large signal
range implies an accurate modelling of the electron density distribution. That
is why we chose to extract fractional charge loss measurements from RC2 data to
verify our particular approach to the electron density distribution modelling
described in Section \ref{sect:density}. 

Thanks to its simple signal shape, it is particularly convenient to study the
charge loss occurring in a charge injection (CI) line. An undamaged CI profile gives a constant signal with a mean value corresponding to the CI
level (cf. Section \ref{sect:tests}), whereas a damaged CI profile typically
shows a strong electron deficit in the leading edge corresponding to
electron captures and a slight electron surplus in the end of the profile as
well as a trailing edge after the CI profile due to the release of electrons from
the traps. To obtain the charge loss one needs to compare the damaged and
undamaged profiles. RC2 data does not contain undamaged CI profile since the
reference data were acquired without CI, it is therefore impossible to know with
any accuracy, the number of electrons injected for a specific injection level. To
compute the charge loss we then assume that the average number of electrons over
the last 4 CI lines (CI duration $= 20$ pixels) would constitute an acceptable
reference. Those last pixels undergo the least charge loss and therefore remain
closest to the actual number of electrons injected per line. The
simulations, for which the true CI level is known, show that this
assumption leads to a slight underestimation of the fractional charge
loss at all signal levels. Hence, the last 4 CI lines are also used in
the simulations to compute the CI reference level and subsequently the
fractional charge loss to avoid any bias in our comparison. The
fractional charge loss, formally the charge loss divided by the charge
injection level times the CI duration, allows us to study the fraction
of signal that is lost due to CTI in a CI profile as a function of the
signal level.

\begin{figure}
  \includegraphics[width=0.49\textwidth]{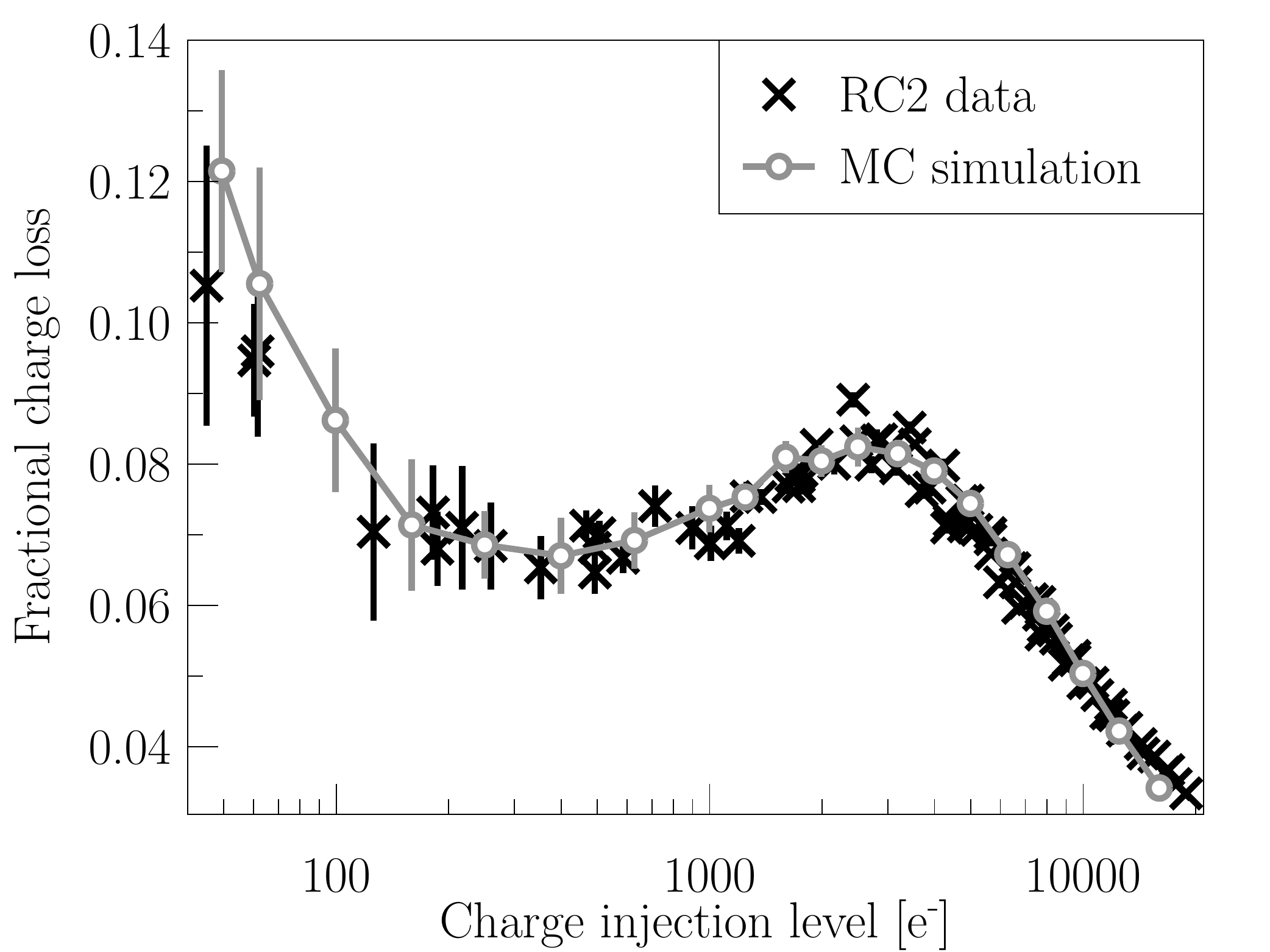}
  \caption{Comparison between the modeled fractional charge loss (grey) and
  experimental data extracted from RC2 (black) as a function of the signal
  level. \label{fig:fcl}}
\end{figure}

In Figure \ref{fig:fcl} the black crosses show the fractional charge loss as a
function of the CI level extracted from RC2 data, we obtain results similar to
\cite{hopkinson2005} (for a direct comparison note that the devices were
irradiated at different doses in the two studies). Fig.\ \ref{fig:fcl} reveals
the complex structure of the pixel by showing a clear break (close to the
nominal SBC full well capacity $1500$ \electron) in the increase of the
fractional charge loss as the signal level diminishes.

To reproduce these measurements with our model we simulate a simplified version
of the experimental setup. For each injection level we simulate the operation of
a CCD with a single pixel column of 4494 pixels, only $2$ scans are performed
(instead of $5$), the CCD contains a unique trap species, no DOB is added, and
the serial register is not simulated. The CCD is operated at the {\gaia}
operational temperature ($163$ K), it is operated in TDI mode as in the RC2 tests. The selected
electron density distribution includes the description of the SBC but no
saturation. A charge injection is performed at the beginning of each scan.
For each considered CI level, we perform this simulation with the same CCD
(i.e. the same CCD characteristics and trap locations), we then measure the
fractional charge loss following the same approach as in the test data analysis.
The whole procedure is repeated $8$ times and the charge loss measurements are
then averaged so that the modeled fractional charge loss is independent from the
noise due to the stochastic nature of the trapping processes.

The model outcomes are then compared to an analytical fit to the experimental
data and $\chi^2$ is used as a comparison criterion.
\begin{equation}
  \chi^{2} = \sum_{i=0}^{S-1} {\frac {\left( \lambda\left(x_{i} \right) -
  \Phi\left(x_{i} \right) \right)^{2}}{\sigma^2}}
\end{equation}
where $\lambda$ is the simulated fractional charge loss, $\Phi$ the fitted fractional charge loss derived from RC2
data, $x_i$ a particular signal level, and $\sigma$ the noise. For a particular
signal level, the experimental data is the mean of $4$ measurements (cf. Section
\ref{sect:tests}).  We assume that the standard deviation from this mean
measurement (black bars in Fig.\ \ref{fig:fcl}) encompasses the experimental
noise, the readout noise as well as the injection noise and is therefore equivalent
to $\sigma$. The set of parameters which minimizes our comparison criterion is
found by using the downhill simplex minimization method \citep{nelder1965}. The
free parameters in our simulation are the trap density, and the density
distribution parameters: the SBC full well capacity, $\eta_{\mathrm{BC}}$, and
$\eta_{\mathrm{SBC}}$. We introduce $\eta$ in order to preserve in the electron density distribution the ratio between the signal confinement volume dimensions $\vect{x}_{max}$. In this way the standard width of the distribution $\sigma_{\vect{x}}$ in each direction corresponds to a fixed fraction of the predefined dimension of the signal confinement volume in that direction $\vect{x}_{max}$:
\begin{equation}
  \sigma_{\vect{x}} = \eta \times \vect{x}_{max}
\end{equation}
To avoid local minima, we first randomly probe the parameter space and establish
a $\chi^2$ map. The downhill simplex is then initialized with the set of
parameters for which the $\chi^2$ is the smallest.
\begin{table}
  \centering 
  \begin{tabular}{|l|l|} 
    \hline
         \multicolumn{2}{|l|}{\textbf{Density parameters} } \\
         \hline
    $\mathrm{FWC}_\mathrm{SBC}\star$ & $2824.89$ \electron \\
    $\mathrm{FWC}_\mathrm{BC}$ & $190\,000$ \electron \\
    $\eta_{\mathrm{BC}}\star$ & $0.101$\\
    $\eta_{\mathrm{SBC}}\star$ & $0.111$\\
    $\sigma_x$ & $1.11\,\mathrm{\mu m}$\\
    $\sigma_y$ & $2.42\,\mathrm{\mu m}$ \\ 
    $\sigma_z$ & $0.076\,\mathrm{\mu m}$ \\
    $\sigma_{y,\mathrm{SBC}}$ & $0.22\,\mathrm{\mu m}$\\
    $\sigma_{z,\mathrm{SBC}}$ & $0.01\,\mathrm{\mu m}$\\
    $y_{0,\mathrm{SBC}}$ & $23\,\mathrm{\mu m}$\\
    $z_{0,\mathrm{SBC}}$ & $0.05\,\mathrm{\mu m}$\\
    \hline
     \multicolumn{2}{|l|}{\textbf{Trap species parameters} } \\
         \hline

    $\rho\star$& $4.08$ traps/pixel \\
    $\sigma$& $5.00\,10^{-20}\;\mathrm{m}^2$\\
    $\tau$&$18.06$ ms\\
    \hline
    \multicolumn{2}{|l|}{\textbf{Goodness-of-fit} } \\
    \hline
    
    $\chi^2$ &$338$\\
    $\chi^2_{\mathrm{red}}$ &$21$\\
    \hline

  \end{tabular} 
  \caption{Simulation parameters and goodness-of-fit for the model result example. The fitted parameters are indicated with a $\star$.} 
  \label{tab:densityBestFit} 
\end{table}
The grey line and circles in Fig.\ \ref{fig:fcl} (left) give an example of the
model results, representative of the quality of fit that can be achieved using
this method. The simulation parameters for this particular fit are summarized in Table \ref{tab:densityBestFit}. The grey bars indicate $1$-sigma deviation from the average over
the 8 simulation iterations. Within the error bars, the
model reproduces the experimental data over a wide range of signal levels
(three orders of magnitude). The CTI mitigation effect of the SBC and the
transition between the BC and SBC signal regimes are particularly well handled.
At higher signal levels the performance of the model slightly degrades. We
notice a deviation that would ultimately lead to underestimate the charge loss
at very high signal levels. It is not possible to confirm this tendency as the
charge injections with a signal level higher than $17$ k\electron could not be
studied due to saturation of the output amplifier. 

A comparison between our analytical description of the electron density distribution and the
more detailed model of a {\gaia} pixel \citep{seabroke2009, seabroke2010} shows that in the serial direction the
Gaussian approximation may be too crude at high signal level, where the distribution saturates quickly and takes a box-like shape.
However, when using SeabrokeÕs model we were not able to reproduce satisfactorily the fractional charge loss measurement. The simulations deviated significantly from the experimental data in particular during the transition between the BC and SBC signal regimes. One possible explanation is that the single e2v Gaia CCD in question has a non-nominal pixel architecture due to manufacturing tolerances (Seabroke's model uses nominal e2v design values).

\subsubsection{Astrometric images} \label{sect:astro}
The verification of the elementary units of the model being satisfactory as well
as the validation of our electron density distribution approach, we are now
interested in estimating the capabilities of the model to reproduce the CTI
induced distortion of stellar images acquired in TDI mode.

Thanks to the second test of RC2 (cf.\ Section \ref{sect:tests}), the model
performance can be investigated for
different stellar brightnesses. To do so we compare the model outcomes with the
stellar images acquired in the irradiated part of the CCD. This comparison is
performed at the sub-pixel level thanks to damaged over-sampled stellar images
built using the multiple scans performed for each set of experimental parameters
(temperature, star brightness, CI delay). The model input signal must be
representative of the CCD illumination conditions. Therefore, we used the accumulated
data in the non-irradiated part of the CCD to create a reference undamaged image
profile. In order to stay as close as possible to the original experimental
set-up, the transit of the investigated star is simulated in two dimensions,
i.e. over a virtual CCD with 12 pixel columns of 4494 pixels. This requires a
two-dimensional input. However during RC2, the images were binned in the serial
direction to recreate in-flight conditions such that the reference curves are LSFs (Line Spread Function, the PSF integrated in the
serial direction). Since we have no information on the two-dimensional PSF, to generate
a PSF from the original reference curve we assume that the profiles in the
parallel and serial directions are the same: $P\left( x,y\right) =
L\left(x\right) \times L\left(y\right)$, where $L$ is the undamaged reference
curve, $x$ and $y$ the positions in pixel respectively in the parallel and serial directions, and $P$ the model input image. The integrated flux of the reference curve
is scaled to produce an input image for each level of illumination (or
artificial star magnitude).

To perform a direct comparison between the model and the over-sampled damaged
profiles, we first extract the sampling scheme specific to each of the damaged
profiles. We then apply it to the PSF generated from the reference curves to
create the model input signal. In this way the required number of simulations
for a single set of parameters to generate an $n$ times over-sampled simulated
damaged profile is $n$. Once the simulations are completed, the individual
predictions are binned in the serial direction and each data point is then
placed at the correct sub-pixel position according to the original sampling
scheme so as to form an oversampled simulated damaged profile.

\begin{figure*}
\includegraphics[width=0.49\textwidth]{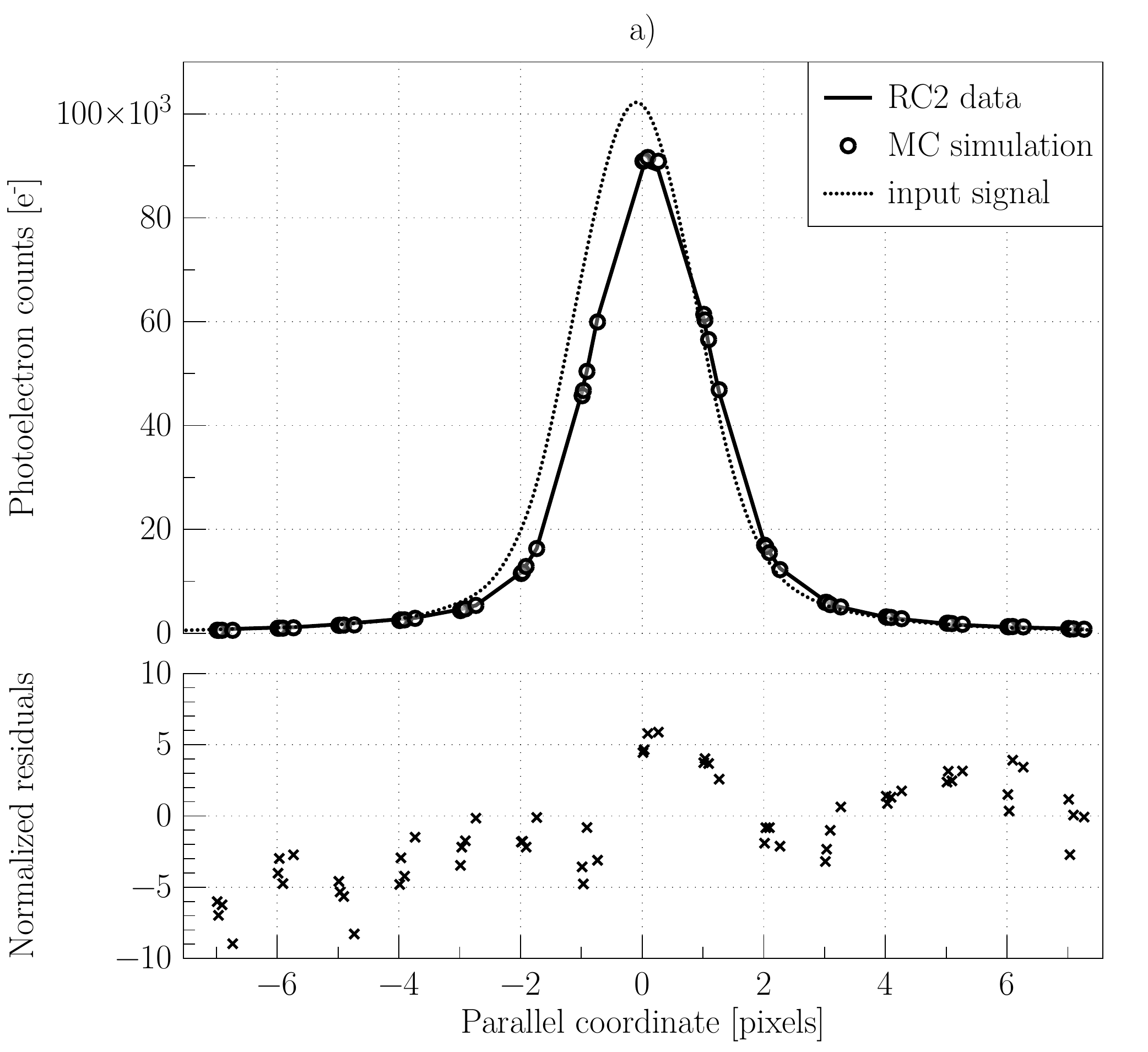}
\includegraphics[width=0.49\textwidth]{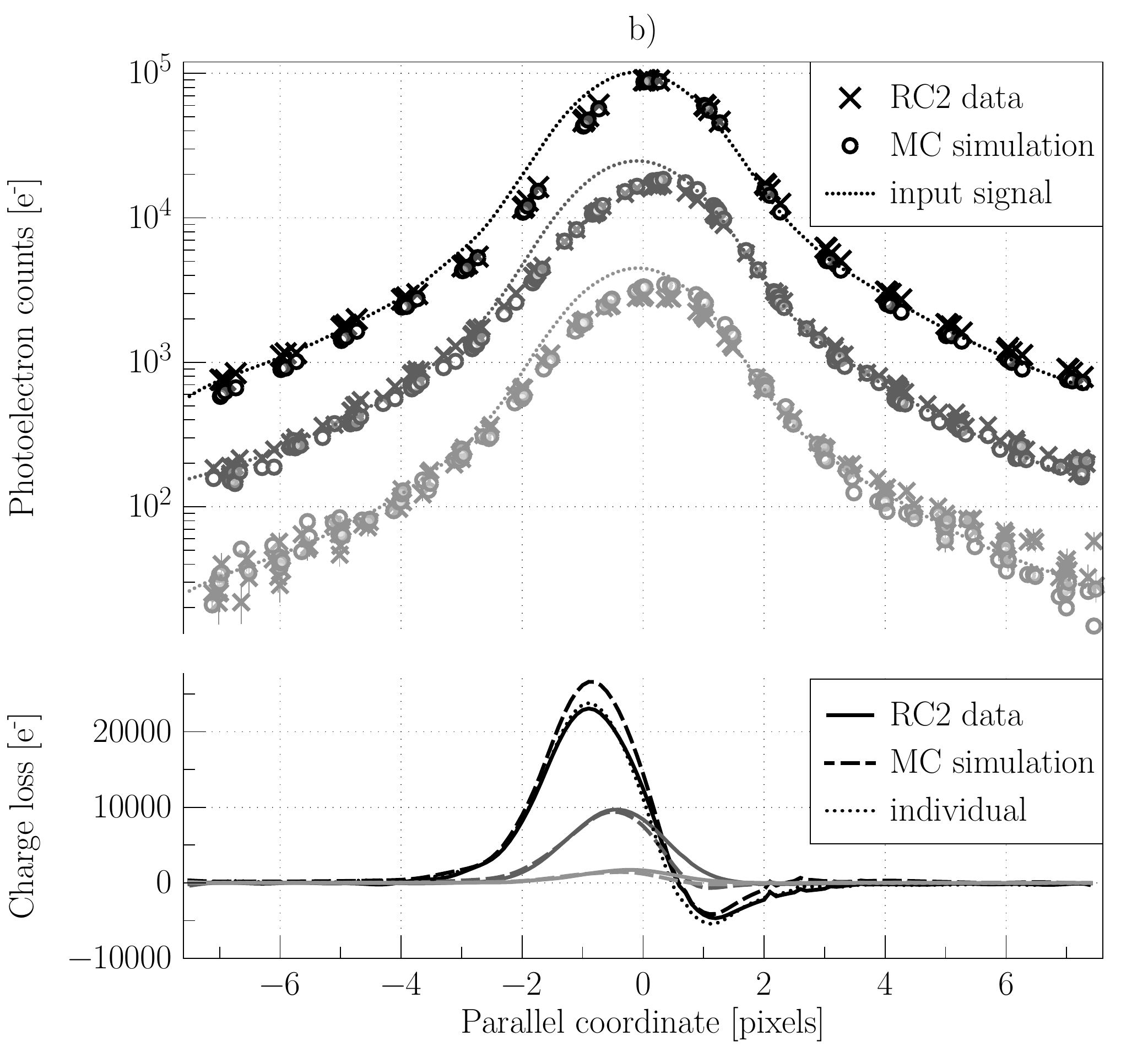}
\caption{(a) \textit{Top left: }Comparison between the RC2 test data (black line) as reduced by \citet{swb2009} and the model simulation (black circles) for a particular signal level and CI delay (Magnitude: $13.63$; CI delay: $\sim30$ s). The charge transfer direction is from right to left. The amplitude of the radiation damage can be appreciated by comparing the damaged profiles (black line and circles) to the input signal (dotted line) based on the non-irradiated test data. \textit{Bottom left: } difference between the simulation and the experimental data normalized by the photon-noise. The quality of fit is similar in the profile core and wings. Note the clustering of the sampling: each data point within a pixel corresponds to a different scan.
\newline(b) \textit{Top right: }Comparison between the RC2 test data (crosses) and the model simulations (circles) for several signal levels and a particular CI delay (Magnitude: $13.63$ (black), $15.17$ (dark grey), $17.03$ (light grey); CI delay: $\sim30$ s). The model simulations are obtained for a single set of parameters (Table \ref{tab:fittingparameters}), which has been optimized to find a reasonable agreement with the test data. Note that this time the ordinate scale is logarithmic in order to facilitate the visual evaluation of the quality of fit for each magnitude and in the core as well as in the wings of the image profiles. The charge transfer direction is from right to left. \textit{Bottom right: } Comparison between the resulting experimental (continuous line) and simulated (dashed line) charge loss as a function of magnitude. The dotted line corresponds to the resulting charge loss for the simulation parameter obtained when fitting to magnitude 13.63 only.}
\label{fig:astrometry}
\end{figure*}

We present the results of this detailed comparison (Fig.\ \ref{fig:astrometry}, Table \ref{tab:fittingparameters}) for two different cases: (i) the model is first used to reproduce a single image profile (i.e. for a unique set of experimental parameters: temperature, CI delay, and magnitude); (ii) and then, with a single set of simulation parameters, the model is used to simultaneously reproduce a set of damaged image profiles with different magnitudes.

When fitting to an individual profile (Fig.\ \ref{fig:astrometry} (a)), $\chi^2$ is used as our
goodness-of-fit criterion:
\begin{equation}
\chi^{2} = \sum_{i=0}^{S-1} {\frac {\left( \lambda\left(x_{i} \right) - N\left(x_{i} \right) \right)^{2}}{\sigma^2}}
\end{equation}
where $\lambda$ is the simulated damaged profile, $N$ the RC2 profile, $x_i$ a particular sub-pixel position, $S$ the total number of data points. $\sigma$, the noise, is considered to be equivalent to the quadratic sum of the photon-noise and the readout noise $r$. The photon-noise is assumed to be Gaussian with a standard deviation of $\sqrt N$ (an approximation of the Poisson statistics for large photon counts) and $r$ is assumed to have the constant value of $4.8$ electrons:
\begin{equation}
  \sigma^2 = N\left(x_i \right) + r^2
\end{equation}
When fitting to a set of damaged profiles (Fig.\ \ref{fig:astrometry} (b)), $\chi^{2}$ is altered in order to avoid any fitting bias towards the brightest magnitudes and the most over-sampled profiles. The new comparison criterion $g$ is thus defined as follows:
\begin{equation}
g = \frac{\chi^{2}}{S \times F} 
\end{equation}
where $F$ is the total integrated flux.

In each case the free parameters of the simulation are the trap parameters ($\rho$ the density, $\sigma$ the capture cross-section, and $\tau$ the release time constant). The number of trap species is fixed to 2. The electron density distribution parameters are fixed to the values summarized in Table \ref{tab:densityBestFit}.
Compared to the fitting of the fractional charge loss measurement, fitting astrometric measurements is a complex task:
\begin{enumerate}
\item The parameter space is larger.
\item The parameter space is more degenerate; a single image profile does not set high constraints on the trap parameters.
\item Both the capture and release processes are involved.
\item The illumination history plays an important role.
\item As we will explain in the following, the number of experimental uncertainties is higher as well as the number of assumptions required by the data processing and the simulation.
\item The fitting also necessitates a high number of two-dimensional simulations of stellar transits, which is computationally very intensive, and sets a high requirement on the minimization procedure efficiency.
\end{enumerate}
The preliminary $\chi^2$ maps also appeared to be hard to characterize as they contained a lot of local minima, none of which presented a satisfactory agreement with the data. In order to better sample the parameter space we decided to perform the first step of our minimization procedure by means of an evolutionary algorithm\footnote[1]{\url{http://watchmaker.uncommons.org/}}. We applied two evolutionary mechanisms, mutation and cross-over, with optimal occurrence probabilities on a restricted initial population of parameter sets and let it evolve towards smaller comparison criteria. After a limited amount of generations, we initialized the downhill simplex algorithm with the set of parameters for which our comparison criterion was the smallest. As previously mentioned, simulating two-dimensional stellar transits is a relatively time-consuming process. For long charge injection delays ($> 20$s) one can neglect the effect of the charge injection release trail, the star then crosses a CCD with almost all the traps empty (from short to fairly long release time-constants). This enables us to drastically reduce the number of simulated TDI steps. Additionally the stellar transits acquired in these conditions offer the advantage of presenting important CTI effects that set a higher constraint on the fitted parameters. We thus selected the subset of RC2 damaged profiles with a CI delay of $\sim30$ s to perform our comparison. Finally, although the profiles were ultimately compared at the sub-pixel level, in order to decrease the total number of simulations, only one sample per pixel was simulated during the fitting procedure
\begin{table}
 \begin{tabular}{|l |l |l |l |}
  \hline
 		\textbf{Fit to}		& \multicolumn{2}{|l|}{single damaged profile} 	 & different signal levels	\\  \hline
    	 	\textbf{Figure}		& \ref{fig:astrometry} a)		& - & \ref{fig:astrometry} b) 		\\ 		\hline
		\textbf{Magnitude}	& 13.67				& 17.03		& 13.67, 15.17, 17.03	\\ 		
		\textbf{CI delay }	& 30s				& 30s		& 30s					\\ 		\hline\hline

		\multicolumn{4}{|l|}{ \textbf{Goodness-of-fit} }			 \\ \hline 
    		$\chi^2_{\mathrm{red}}$		& $15.4$							& $14.4$					& $70.0$,  $41.4$, $18.2$			\\  \hline \hline
		\multicolumn{4}{|l|}{ \textbf{Trap parameters} }		\\ \hline 									 
		$\rho_1$  [per pixel] 	& $0.57$						& $4.59$					& $3.97$					\\
		$\sigma_1$ [m$^{2}$]		& $6.56\,10^{-21}$			& $3.99 \, 10^{-21}$		& $2.33 \, 10^{-21}$	\\
		$\tau_1$ [s]             & $2.23\, 10^{-3}$			& $0.60$					& $0.56$ 	\\ \hline
		$\rho_2$   				& $4.11$						& $4.39$					& $4.10$					\\
		$\sigma_2$       		& $1.17\, 10^{-20}$			& $1.89 \, 10^{-22}$		& $2.38 \, 10^{-22}$	\\
		$\tau_2$               	& $97.34$					& $2.92$					& $78.91$					\\ \hline
	 \end{tabular}
	 \caption{MC model fitting parameters corresponding to the examples shown in Figs.\ \ref{fig:astrometry} a) and b). Note that $\chi^2_{\mathrm{red}}$, the reduced $\chi^2$, indicates here the goodness-of-fit. In the three presented cases, a short and a long release time constant species are needed to explain the experimental data.}
 \label{tab:fittingparameters}
\end{table}
\begin{figure*}
\includegraphics[width=0.98\textwidth]{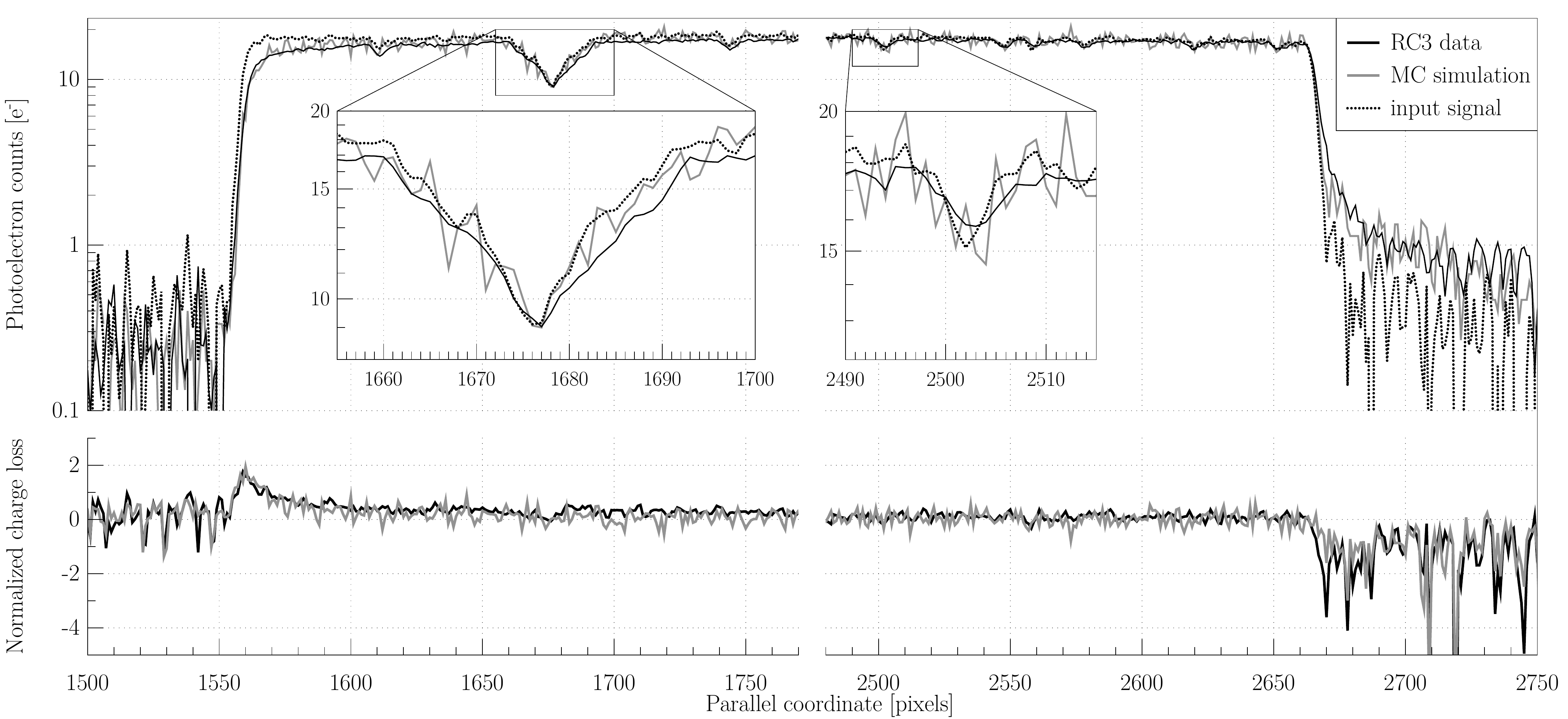}
\caption{\textit{Top: }Comparison between the RC3 test data (black line) as reduced by \citet{hall2010} and the model simulation (grey line) for a star of magnitude $12.5$ (No CI, and lowest level of background light ($< 1$ \electron/pixel)). The charge transfer direction is from right to left. The reference averaged spectrum (acquired in the non-irradiated part of the CCD) was used as input signal (dotted).
\textit{Bottom: } The resulting charge loss normalized by the photon-noise for the simulation (grey) and the test data (black).}
\label{fig:rvs}
\end{figure*}
Figure \ref{fig:astrometry} (a) presents an example of a simulated damaged profile (circles), this profile is representative of the best-fitting achievable in these particular conditions at a magnitude of $13.67$ and should be compared to the black line profile (RC2 data). As can be seen from the residuals normalized by the photon-noise, the simulation is in fairly good agreement with the data over the whole profile. However we note a slightly larger disagreement in the image leading edge. This is not surprising as for bright images the wings still contain a fairly large amount of electrons (e.g., 800 \electron in the first simulated sample of the leading edge), which may play an important role in the self-illumination history by filling a number of traps and mitigating the CTI effects in the image core. Hence the fit may be further improved by simulating a profile wider than 15 pixels. The best individual fits obtained at fainter magnitudes (cf. example in Table \ref{tab:fittingparameters}) shows a slightly better overall agreement. The relative amplitude of the CTI effects in the core compare to the wings is very sensitive to the input image shape in the CCD serial direction. The assumption made in order to build a two-dimensional input signal may thus also limit the ultimate goodness-of-fit.

Figure \ref{fig:astrometry} (b) shows the resulting fits at three different magnitudes ($13.67$, $15.17$, $17.03$) obtained for a single set of simulation parameters (cf. Table \ref{tab:fittingparameters}). The overall agreement is remarkable. As can be seen in the bottom part of the plot, the amplitude of the CTI effects is qualitatively well reproduced as a function of the signal brightness. And as expected the charge loss peak slowly shifts towards the image core with fainter signal. However one should notice that when a single set of trap parameters is fitted to several profiles at different signal levels the resulting goodness-of-fit for each individual profile decreases, see Table \ref{tab:fittingparameters}. This is also illustrated in the bottom part of Fig.\ \ref{fig:astrometry} (b), where the resulting charge loss curves at the magnitude $13.67$ for the individual fit case (black dotted line) can be compared to the multiple magnitudes fit case (black dashed line). This difference in goodness-of-fit may be ascribed to variations in the experimental conditions between two tests with different signal level (slightly different temperatures and illumination background). Also the intervals between the tests were several days. In between two tests the CCD was stored at room temperature. Astrium later acknowledged that this may have resulted in a change of the CCD state, as discrepancies in test results were observed for similar experimental conditions. As already discussed in \cite{prodhomme2010}, this may set a limit on the ultimate goodness-of-fit achievable with a single set of trap parameters to several profiles by any model. As a final remark on this part of the comparison, we would like to state that even if the trap species parameters found as a result of the fitting procedure could be associated with known trap species from the literature, the uncertainties in the experiments are too large and the assumptions in the simulation process too numerous to conclude that such species were indeed present in the tested CCD. If one wants to infer trap species parameters by using this model to reproduce test data, the experiment should be carefully designed to serve this purpose. Artificial charge injections would be particularly suited for such an experiment as one can infer trap densities and capture cross-sections from the charge loss that occurs in their profile and release time constants from their trailing edge. We would in particular recommend: (i) to repeat the tests at different temperatures in order to break the potential degeneracies in the time release constant space; (ii) to use different levels of charge injection from a few electrons to the pixel full well capacity, in order to make sure that no trap is missed as well as to set the highest constraint possible on the charge density distribution model over a complete signal range. 

\subsubsection{Radial velocity spectrometer images} \label{sect:rvs}
The radial velocity of stars will be measured on-board {\gaia} thanks to a medium resolution spectrometer \citep{katz2004} that will enable the analysis of the Doppler shift of the stellar spectral lines, in particular the Ca II triplet. The spectrometer is composed of a series of prisms and a diffraction grating that will disperse the stellar light before it reaches the red-enhanced {\gaia} CCDs. The combined effect of TDI and high light dispersion will result in a very faint spectral signal, down to a few electrons per spectral feature. The signal-to-noise ratio for a single measurement is expected to be very low (e.g., $7$ for a star of magnitude $16$), and, at the faint end, the radial velocity measurement will rely on the co-addition of multiple spectra to recover the spectral features. Until recently CTI had never been studied at such low signal levels, and it was uncertain if one could indeed recover spectral features by co-adding damaged spectra. RC3 has been designed to address this particular issue by the use of a mask manufactured to reproduce in detail the spectrum of a G2V star. To illustrate the capacity of our model to reproduce the CTI effects in TDI mode at extremely low signal levels ($< 20$ \electron) we use the RC3 test data and compare the model outcomes to the stellar spectra acquired in the irradiated part of the CCD.

The model input signal is built from the accumulated data in the non-irradiated part of the CCD and the transits are simulated over a virtual CCD containing 6 pixel columns of 4494 pixels. During the tests the spectra were binned over 12 pixels in the serial direction, we thus again had to assume the serial signal profile to perform two-dimensional simulations. We assumed the same parallel spectral profile in every pixel column but with different intensities. The fractions of the total intensity for each pixel column were chosen according to the serial profile of the simulated spectrum provided by the Gaia Instrument and Basic Image Simulator, GIBIS (0.04, 0.14, 0.26, 0.29, 0.20, 0.07). This time we do not perform any optimization of the trap parameters but we selected two trap species from \cite{seabroke2008} (divacancy and unknown: $\rho_{1,2}=$ 1 trap/pixel, $\sigma_{1,2} =5\,10^{-20}\,\mathrm{m^2}$, $\tau_1=$ 20 s, and $\tau_2=$ 80 ms). Once again we fixed the electron density distribution parameters to the values summarized in Table \ref{tab:densityBestFit}.

In these conditions, one can obtain a spectrum such as the one presented in Fig.\ \ref{fig:rvs} for a star of magnitude $12.5$ (grey line). In a similar fashion to the depicted experimental damage spectrum (black line), 15 transits were averaged to obtain the simulated spectrum. In both the RC3 data and the model outcome, one can notice that even at such low signal level the CTI effects do not completely wash away the spectral features; they can be recovered by co-adding several spectra. However, as can be seen in more detail from the blown-up portions of the presented spectra, CTI lowers the continuum and affects the absorption features by increasing their width and shifting them. These effects lead to a significant decrease of the overall spectral contrast. In the bottom part of the figure, we show the charge loss normalized by the photon-noise. The model reproduces remarkably well the first significant charge loss bump (only few electrons) occurring at the leading edge of the spectrum (left). The electron release that occurs at the trailing edge after the spectrum transit (right) is also well reproduced. The simulated averaged profile is binned in the serial direction over 6 pixels instead of 12 during the tests, as a consequence the simulation is slightly noisier than the RC3 data. It is thus hard to distinguish, but within the spectrum itself one can notice that the model does widen and shift the spectral features (see the blow-ups). The amplitude of the shift and increase in width is not quite matched though, this could be imputed to a difference in the release time constant for the trap species.

\section{Conclusion}\label{sect:conclusion}
We have described a physical Monte Carlo model that simulates CTI effects
induced by radiation damage in astronomical CCDs at the electrode level. This
model has been elaborated in the challenging context of ESA's {\gaia} mission. The operating conditions, the extreme astrometric requirements and a novel CCD pixel architecture,
necessitated the development of a new approach in the representation of the
electron density distribution as well as a more detailed description of the
trapping probabilities. These new features have been combined in a comprehensive
simulation of CCD charge collection and transfer at the electrode level. In
order to verify the model predictions, we first validated the unit blocks of the
simulation and then proceeded to a detailed comparison between the model and experimental test data. We showed that the model is able to
accurately reproduce the CTI effects for a wide range of signal levels down to
a few electrons, hence validating our electron density distribution
representation. We finally validated the global model operation by assessing the
model capability to reproduce the CTI induced distortion on stellar images and spectra
acquired in TDI mode at different magnitudes. 

The model elaboration contributed greatly to our present understanding of
CTI effects. In particular, by showing that the implementation of a density driven approach of the electron packet growth enables the reproduction of experimental data and that to be successful in modelling the CTI effects at very faint signal levels, no detail
should be neglected. The simulations should be as realistic as possible, down
to the transfer of electrons at the electrode level and the simulation of each
individual trap.

Due to the complexity of the CTI effects (including the varying illumination
history) and the extreme accuracy required, in particular on the estimation of
the stellar image location, one cannot apply a conventional correction of the
raw data. Even if this were possible, the process would likely be unstable and
lead to poorly understood error propagation. Instead of a direct correction,
the {\gaia} Data Processing and Analysis Consortium (DPAC) is developing a
scheme which relies on a forward modelling approach that enables the
estimation of the true image parameters from a damaged observation
\citep[e.g.,][]{prodhomme2010b}. In this scheme each observation is ultimately
compared to a modelled charge profile in which the distortion of the CTI-free
image (PSF) will be simulated through an analytical CTI model. Currently the
Monte Carlo model described in this paper is used to generate large synthetic
data sets of both damaged and damage-free observations. The simulations are
used to re-assess the final performance of the mission, taking the effects of
radiation damage into account, as well as to verify the DPAC CTI mitigation
scheme \citep{prodhomme2011, holl2011}.

Although developed in the {\gaia} context, we tried to keep the model as general
and as flexible as possible. It can be used to simulate any kind of measurements
performed with a CCD operated in imaging mode or TDI mode. Different clocking schemes
can be applied and the description of the electron density distribution should
be flexible enough to simulate different pixel architectures. The use of this
model is particularly relevant in the frame of space experiments that aim at
very accurate measurements at low signal levels. ESA's \textit{Euclid}
mission, the astrometric measurements performed on board {\hst} or future
X-ray missions sent to L2 may benefit from the use of such a model
to evaluate the impact of radiation damage on their performance budgets.

A running version of the model as well as a brief documentation and a few examples
are readily available at \url{http://www.strw.leidenuniv.nl/~prodhomme/cemga.php} as
part of the {\sc CEMGA} java package (CTI Effects Models for {\gaia}) developed at
Leiden Observatory. Please contact the authors for more information on how to
use the package.

\section*{Acknowledgments}

The authors would like to acknowledge EADS Astrium for kindly supplying the experimental test data necessary to accomplish the verification of our model.
We would like to thank F. van Leeuwen from the Institute of Astronomy in Cambridge, and D.J. Hall from the e2v centre for electronic imaging at the Open University for providing the results of their reduction and analysis of the test data to support the model verification effort.
This model and study greatly benefited from the numerous discussions with all the participants of the {\gaia} Radiation Task Force in particular M.\ Weiler (Observatoire de Paris-Meudon), B.\ Holl (Lund Observatory), G.M.\ Seabroke (MSSL), R.\ Kholey (ESA-ESAC), and F.\ Raison (ESA-ESAC). This work has been funded by the European Community's sixth framework programme (FP6) through the Marie Curie research training network ELSA (European Leadership in Space Astrometry) contract No MRTN-CT-2006-033481.

\bibliographystyle{mn2e}
\bibliography{references.bib}

\label{lastpage}

\end{document}